\documentclass[12pt,draftclsnofoot,onecolumn]{IEEEtran}

\usepackage{color}
\usepackage{multirow}
\usepackage{graphicx}
\usepackage{epstopdf}
\usepackage[cmex10]{amsmath}
\usepackage{amssymb}

\usepackage{multirow}
\usepackage{amsthm}
\usepackage{cite}
\usepackage{flushend}
\usepackage{acronym} 
\usepackage{algorithm, algcompatible}

\newcommand{\diag}{\mathop{\rm diag}}
\newcommand{\Ei}{\mathop{\rm Ei}}

\newtheorem*{lemma}{Lemma}

\newtheorem{corollary}{Corollary}
\newtheorem{proposition}{Proposition}

\algnewcommand\INPUT{\item[\textbf{Input:}]}
\algnewcommand\OUTPUT{\item[\textbf{Output:}]}

\allowdisplaybreaks

\makeatletter
\newcommand{\doublewidetilde}[1]{{%
  \mathpalette\double@widetilde{#1}%
}}
\newcommand{\double@widetilde}[2]{%
  \sbox\z@{$\m@th#1\widetilde{#2}$}%
  \ht\z@=.9\ht\z@
  \widetilde{\box\z@}%
}
\makeatother

\begin{document}

\title{MIMO Underlay Cognitive Radio:\\Optimized Power Allocation, Effective Number of Transmit Antennas and Harvest-Transmit Tradeoff}

\author{Nikolaos~I.~Miridakis, Theodoros~A.~Tsiftsis,~\IEEEmembership{Senior Member,~IEEE}\\ and George C. Alexandropoulos~\IEEEmembership{Senior Member,~IEEE}
\thanks{N. I. Miridakis is with the Department of Computer Systems Engineering, Piraeus University of Applied Sciences, 12244, Aegaleo, Greece (e-mail: nikozm@unipi.gr).}
\thanks{T. A. Tsiftsis is with the School of Engineering, Nazarbayev University, Astana 010000, Kazakhstan (e-mail: theodoros.tsiftsis@nu.edu.kz).}
\thanks{G. C. Alexandropoulos is with the Mathematical and Algorithmic Sciences Lab, Paris Research Center, Huawei Technologies
France SASU, Boulogne-Billancourt 92100, France (e-mail: george.alexandropoulos@huawei.com).}
}


\maketitle

\begin{abstract}
In this paper, the performance of an underlay multiple-input multiple-output (MIMO) cognitive radio system is analytically studied. In particular, the secondary transmitter operates in a spatial multiplexing transmission mode, while a zero-forcing detector is employed at the secondary receiver. Additionally, the secondary system is interfered by single-antenna primary users (PUs). To enhance the performance of secondary transmission, optimal power allocation is performed at the secondary transmitter with a constraint on the maximum allowable outage threshold specified by the PUs. Further, the effective number of secondary transmit antennas is specified based on the optimal power allocation for an arbitrary MIMO scale. Also, a lower bound on the ergodic channel capacity of the secondary system is derived in a closed-form expression. Afterwards, the scenario of a massive MIMO secondary system is thoroughly analyzed and evaluated, where the harvesting-enabled secondary transmission is studied. The optimal power allocation, the effective number of secondary transmit antennas, the efficient tradeoff between transmit-and-harvest secondary antennas, and the average channel capacity of the secondary system are analytically presented. Finally, extensive numerical and simulation results corroborate the effectiveness of our analysis, while some useful engineering insights are provided. 
\end{abstract}


\IEEEpeerreviewmaketitle

\section{Introduction}
\IEEEPARstart{C}{ognitive} radio (CR) is widely recognized as a promising technique to resolve the issue of spectrum scarcity, caused by the explosive growth of wireless data traffic. Among various paradigms to deploy CR in practice, underlay CR allows simultaneous transmissions of primary users (PUs) and secondary users (SUs), as well as low implementation complexity \cite{j:TanabHamouda2017}. In a practical underlay CR system, to guarantee the quality of service (QoS) of PUs granted spectrum resources, the transmit (Tx) power of SUs with no fixed spectrum resources is strictly limited, such that the harmful interference from SUs to PUs remains below a prescribed tolerable level. To improve the performance of secondary transmission, multiple-input multiple-output (MIMO) and even massive MIMO antenna techniques can be explored since they provide additional degrees of freedom (DoF) in spatial domain, compared with traditional single-antenna transmission \cite{j:MiridakisTsiftsisAlexandropoulos2017}. When MIMO antenna was integrated into underlay CR systems, the spatial diversity gain of MIMO was widely used to enhance the reliability of secondary transmission, see e.g., \cite{j:Sarvendranath2013, c:XiongMukherjee2015} and references therein. On the other hand, the spatial multiplexing gain of MIMO was exploited to improve the data rate of secondary transmission, see e.g., \cite{c:YangQaraqe2013}. 

A fundamental challenge for the modeling of practical underlay CR systems is how to mitigate the detrimental effect of imperfect channel state information (CSI). Most existing literature works thus far proposed the underlay CR transmission under the so-called interference temperature (IT) factor \cite{j:SharmaBogale2015}; i.e, the secondary Tx power should not exceed a certain threshold so as not to cause unexpected (harmful) interference onto the primary system. Yet, in order to provide the required transparency across the primary service and, thereby, an efficient secondary communication, the transmission under the IT constraint implies a perfect CSI knowledge between the primary and secondary channel gains. However, in practice, the secondary system cannot acquire perfect CSI with respect to the primary nodes (e.g., due to user mobility and/or the lack of sufficient feedback signaling from the primary nodes). According to the seminal work in \cite{j:ChenSi2012}, the probability that the instantaneous secondary interference at the primary receiver (Rx) exceeding IT \emph{is always $50\%$ in the case of imperfect CSI, regardless of the CSI imperfection} \cite[Eq. (4)]{j:ChenSi2012}. Therefore, the IT-enabled approach becomes a rather inefficient mode of operation. Another alternative is to implement secondary communication using average (statistical) CSI between the secondary-to-primary channel links, which represents a much more feasible solution \cite{j:XuZhao2015}. In this case, the secondary Tx power should not exceed a corresponding average IT threshold. Nonetheless, by adopting this approach, the resultant instantaneous secondary Tx power can produce certain `peak rates', which in turn may cause unexpected (instantaneous) interference onto the primary Rx \cite{j:MiridakisXia2017}. 

A more robust transmission strategy is to constrain the secondary Tx power subject to a \emph{probabilistic} approach with respect to the total signal-interference-plus-noise ratio (SINR) at the primary Rx \cite{j:MusavianAissa2009,j:SmithDmochowski2013}. Such a constraint guarantees that the received SINR at the primary system stays below a tolerable level within a prescribed probability. This metric is more resilient to the unwanted (yet realistic) scenario of imperfect CSI between the two heterogeneous systems, since it ensures a predefined QoS of the primary communication for a given probability, whereas statistical-only CSI knowledge is required. 

On another front, the massive MIMO transmission is emerging at the forefront of wireless communications research nowadays. Its great potential can find direct application in underlay CR systems due to the enormously high channel gains and low Tx power regimes that can provide. This is achieved by an increased spatial DoF provided by massive MIMO, which are (at least) one order of magnitude higher than the conventional MIMO antenna arrays. Most importantly, such an increased DoF may even enhance the energy efficiency, when energy harvesting is implemented at the considered system nodes \cite{j:MishraAlexandropoulos2017}. Under this regime, a massive MIMO transmitter may harvest energy via its vast antenna array, which in turn can be used to transmit its data afterwards, accordingly. In the case of underlay CR systems, the latter effect can be much more impactful since the corresponding CR nodes may benefit from the presence of primary transmissions so as to enhance their energy harvesting activity \cite{j:PrasadHosain2017,j:HuangHan2015}.

In current work, the performance of MIMO underlay CR systems is studied from the green communications perspective. More specifically, the Tx power allocation of the secondary system is formulated, by satisfying a prescribed QoS level for the primary service, while optimizing the secondary communication at the same time. A probabilistic approach based on the total received SINR at the primary system is adopted, which is an efficient solution for the practical scenario of statistical-only CSI knowledge between the channel gains of the two heterogeneous systems. To further enhance the data rate of the secondary system, the spatial multiplexing transmission mode is adopted, where multiple independent streams are simultaneously transmitted at the secondary transmitter (ST). At the secondary receiver (SR), the computationally efficient linear zero-forcing (ZF) detection is applied, where perfect CSI between the $\rm ST - SR$ links is available. In addition, the effective number of secondary Tx antennas is derived via a cost-effective iterative algorithm. The motivation behind the latter strategy is that reducing the number of (uncoded) secondary Tx antennas, the total interference reflected at the primary Rx is being proportionally reduced. At the same time, the remaining active Tx antennas can be used to satisfy a given QoS for the secondary communication. The presented analytical results rely on mutually independent Rayleigh faded channels. Also, the analysis considers the general case of arbitrary number of antenna arrays. For the special scenario of massive MIMO, the concept of energy harvesting-enabled secondary transmission is introduced, where the secondary Tx benefits from quite a high spatial DoF. In this case, the (potentially many) inactive secondary Tx antennas can harvest energy from primary transmissions and also from the remaining active (nearby) secondary Tx antennas. Hence, the optimal Tx power allocation problem and the effective number of active/inactive Tx antennas are both revisited, while some useful engineering outcomes are manifested.

At this point, it is worthy to state that the joint problem of Tx power allocation and the effective number of Tx antennas regarding spatial multiplexed MIMO underlay CR systems has not been studied elsewhere in the open technical literature. More so, the special case of massive MIMO with energy harvesting capabilities has not been studied thus far. The main contributions of this work are summarized as follows:

\begin{itemize}
	\item A new simple power allocation scheme is designed for ST in underlay MIMO CR systems.
	\item Based on the latter scheme, the effective number of secondary Tx antennas is provided with the aid of an iterative antenna reduction algorithm.
	\item A closed-form lower bound on the ergodic channel capacity of the secondary system is derived.
	\item The aforementioned Tx power allocation and the effective number of Tx antennas at the secondary system are further specified for the special case of secondary massive MIMO systems, when energy harvesting is performed at ST.
	\item To approach realistic energy harvesting conditions, the \emph{piece-wise} linear model in \cite{j:DongHossain2016} is considered, which reflects more accurately the performance of a practical energy harvesting circuit.
	\item Based on the derived analytical results, the average channel capacity of the secondary massive MIMO system is obtained in a closed-form expression. 
\end{itemize}

To detail the aforementioned contributions, the rest of this paper is organized as follows. Section \ref{System Model} describes the considered system model. Section \ref{Performance of Secondary System} devises an optimal power allocation at ST, provides the effective number of secondary Tx antennas and presents a lower bound of the ergodic channel capacity in a closed form. Section \ref{Massive MIMO Energy Harvesting-Enabled Secondary System} analyzes the performance of the considered system in the massive MIMO scale, while the scenario of energy harvesting at the secondary system is thoroughly analyzed. Afterwards, Section \ref{Numerical Results} presents simulation results compared with numerical ones, while Section \ref{Conclusions} concludes the paper. Some detailed derivations are relegated to Appendix.

{\bf Notation}: Vectors and matrices are denoted by lowercase and uppercase bold symbols (e.g., $\mathbf{x}$ and $\mathbf{X}$), respectively. The superscripts $(\cdot)^{-1}$,  $(\cdot)^{\dagger}$ and $(\cdot)^{\mathcal{H}}$ means the inverse, pseudo-inverse and conjugate transpose, respectively. $\mathbf{x}_{i}$ denotes the $i^{\rm th}$ entry of $\mathbf{x}$ while $[\mathbf{X}]_{ij}$ stands for the $(i, j)$ element of $\mathbf{X}$. $\diag\{x_{i}\}^{n}_{i=1}$ means a diagonal matrix with entries $x_{1}, \cdots, x_{n}$.  The operator $(x)^{+}$ equals $x$ if $x>0$, and zero otherwise. $|x|$ takes the absolute value of $x$ while $\|\mathbf{x}\|$ is the Euclidean norm of $\mathbf{x}$. $\mathbf{I}_{v}$ stands for the identity matrix of size $v \times v$. $\mathbb{E}[\cdot]$ is the expectation operator and ${\rm Pr}[\cdot]$ returns probability. The symbol $\overset{\text{d}}=$ means equality in distribution. The functions $f_{X}(\cdot)$, $F_{X}(\cdot)$ and $\mathcal{M}_{X}(\cdot)$ represent probability density function (PDF), cumulative distribution function (CDF) and moment generating function (MGF) of a random variable (RV) $X$, respectively. Complex-valued Gaussian RVs with mean $\mu$ and variance $\sigma^{2}$ is denoted as $\mathcal{CN}(\mu,\sigma^{2})$ while central chi-squared RVs with $v$ DoF as $\mathcal{X}^{2}_{v}$. Also, $\mathcal{L}^{-1}\{\cdot\}$ denotes the inverse Laplace transform. Moreover, $\mathcal{O}(\cdot)$ represents the Landau symbol. Finally, $\Gamma(\cdot)$ denotes the Gamma function \cite[Eq. (8.310.1)]{tables}, $\Gamma(\cdot,\cdot)$ is the upper incomplete Gamma function \cite[Eq. (8.350.2)]{tables}, $\psi(\cdot)$ is the digamma function \cite[Eq. (8.360.1)]{tables}, $\Ei(\cdot)$ represents the exponential integral function $\Ei$ \cite[Eq. (8.211.1)]{tables}, ${\rm E}_{1}(\cdot)$ represents the first-order exponential integral function \cite[p. xxxv]{tables}, $I_{0}(\cdot)$ is the zero-order modified Bessel function of the first kind \cite[Eq. (8.445)]{tables}, and $\Phi_{2}(\cdot,\cdot;\cdot;\cdot,\cdot)$ denotes the Humbert confluent hypergeometric series $\Phi_{2}$ \cite[Eq. (9.261.2)]{tables}.

\section{System Model}
\label{System Model}
\begin{figure}[!t]
\centering
\includegraphics[keepaspectratio,width=2.8in]{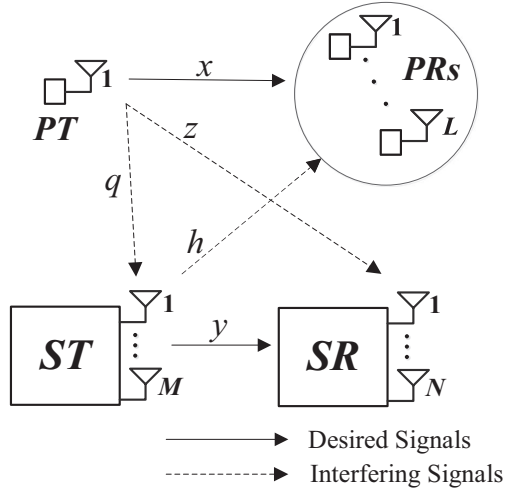}
\caption{The considered system configuration. The parameters $x$, $y$, $h$, $q$ and $z$ denote the involved channel gains between the primary and secondary system, which are explicitly defined hereinafter.}
\label{fig1}
\end{figure}

Consider a point-to-point underlay (secondary) MIMO CR system, which operates under the presence of a primary system, as illustrated in Fig.~\ref{fig1}, where PT and PR denote the primary Tx and Rx, respectively. In particular, ST and SR are equipped with $M$ and $N\geq M$ antennas, respectively. Also, a single-antenna primary system is assumed, which consists of a base station and $L$ corresponding primary nodes/users. The secondary Tx antennas operate in a spatial multiplexing mode and $M$ independent data streams are simultaneously transmitted in a given time instance. ZF detection is adopted at the receiver side. Independent Rayleigh channel fading conditions are assumed for all the involved links. Moreover, both ST and SR are able to acquire statistical (second-order) CSI with respect to the channel gains of the primary system,\footnote{In principle, CSI of the links between the primary and secondary nodes can be obtained through a feedback channel from the primary service or via a band manager that mediates the exchange of information between the primary and secondary networks \cite{j:MiridakisXia2017}.} while
perfect CSI is assumed regarding the channel gains between ST and SR.

Upon the reception of secondary streams, the received signal at SR is given by
\begin{equation}
\mathbf{r} 
= \mathbf{H}\mathbf{P}^{\frac{1}{2}}\mathbf{s}+\sqrt{p_{\rm p}}\mathbf{h}_{\text{p}}s_{\text{p}}+\mathbf{w},
\label{eq1}
\end{equation}
where $\mathbf{H}\in \mathbb{C}^{N\times M}$ denotes the desired channel from ST to SR, $\mathbf{s} \in \mathbb{C}^{M\times 1}$ represents the transmitted signals from the secondary source, $\mathbf{h}_{\text{p}}\in \mathbb{C}^{N\times 1}$ stands for the interfering channel from PT to SR, $s_{\text{p}} \in \mathbb{C}^{1\times 1}$ is the transmitted signal from PT, and $\mathbf{w} \in \mathbb{C}^{N\times 1}$ models the additive white Gaussian noise (AWGN) at SR. Moreover, $\mathbf{P} \in \mathbb{R}^{M\times M} = \diag\{p_{i}\}^{M}_{i=1}$ is a diagonal matrix with $p_i$ being the optimal Tx power at the $i^{\rm th}$ antenna of ST (to be explicitly determined in Section~\ref{Performance of Secondary System}), $p_{\rm p}$ is a constant used to denote the fixed Tx power at PT, and $\mathbf{w}\overset{\rm d} = \mathcal{CN}(\mathbf{0}, N_{0}\mathbf{I}_{N})$ with $N_{0}$ being the AWGN variance. Without loss of generality, the signals at either ST or PT are normalized, i.e., $\mathbb{E}[\mathbf{s}\mathbf{s}^{\mathcal{H}}] = \mathbf{I}_{M}$ and $\mathbb{E}[s_{\text{p}}s_{\text{p}}^{\mathcal{H}}] = 1$.

Based on the principle of ZF detection, the detected symbol vector can be written as
\begin{equation}
\hat{\mathbf{r}} 
 \triangleq \mathbf{G}^{\dagger} \mathbf{r}
 = \mathbf{s}+\sqrt{p_{\rm p}}\mathbf{G}^{\dagger}\mathbf{h}_{\text{p}}s_{\text{p}}+\mathbf{G}^{\dagger}\mathbf{w},
\label{rzf}
\end{equation}
where 
\begin{equation}
\mathbf{G}^{\dagger} 
\triangleq (\mathbf{H}\mathbf{P}^{\frac{1}{2}})^{\dagger}
 = \left((\mathbf{H}\mathbf{P}^{\frac{1}{2}})^{\mathcal{H}}\mathbf{H}\mathbf{P}^{\frac{1}{2}}\right)^{-1}(\mathbf{H}\mathbf{P}^{\frac{1}{2}})^{\mathcal{H}}.
\end{equation}
The above system model of \eqref{rzf} has been well-investigated in the bibliography so far in terms of performance analysis. The resultant SINR of the $i^{\rm th}$ secondary stream ($1\leq i \leq M$) is distributed as \cite{j:MatthaiouZFMassiveMIMO2013}
\begin{align}
{\rm SINR}_{i}\overset{\text{d}}=\frac{p_{i} y_{i}}{p_{\rm p}z+N_{0}},
\label{sinrdistr}
\end{align}
where
\begin{align}
f_{y_{i}}(y)=\frac{y^{N-M}\exp(-y/\mathbb{E}[y])}{\mathbb{E}[y]^{N-M+1}(N-M)!},
\label{ydistr}
\end{align}
and
\begin{align}
f_{z}(z)=\frac{\exp(-z/\mathbb{E}[z])}{\mathbb{E}[z]},
\label{zdistr}
\end{align}
while $\mathbb{E}[y]$ and $\mathbb{E}[z]$ stand for the average channel gains between the ${\rm ST}-{\rm SR}$ and ${\rm PT}-{\rm SR}$ links, respectively.

On the other hand, the SINR at the $j^{\rm th}$ PR ($1\leq j\leq L$) is distributed as
\begin{align}
{\rm SINR}^{(pr.)}_{j}\overset{\text{d}}=\frac{p_{\rm p} x_{j}}{\sum^{M}_{i=1}p_{i} h^{(j)}_{i}+N_{0}},
\label{sinrprimary}
\end{align}
where $x_{j}$ denotes the effective channel gain of the ${\rm PT - PR}_{j}$ link, which is exponentially distributed and has an average channel gain $\mathbb{E}[x]$. Also, $h^{(j)}_{i}$ stands for the interfering power (exponentially distributed with average power $\mathbb{E}[h^{(j)}]$) between the ${\rm ST}_{i}{\rm - PR}_{j}$ link.

\section{Performance of Secondary System}
\label{Performance of Secondary System}
We commence by formulating the considered optimization problem. Then, the optimal power allocation and the effective number of secondary Tx antennas are presented.

\subsection{Optimized power allocation and effective $M$}
Since multiple secondary Tx antennas operate in the spatial multiplexing transmission mode, maximizing the data rate of secondary transmission is equivalent to maximizing the achievable data rate of all the involved secondary data streams. In turn, this can be achieved by proportionally maximizing the corresponding Tx power at each antenna. Yet, this Tx power should be bounded, in order not to cause any unexpected harmful interference to PR. Additionally, allowing more or less active secondary Tx antennas directly influences the transmit power allocation and the total power budget. Thereby, the considered joint optimization problem reads as
\begin{subequations}
\begin{align}
\mathcal{P}_{1}: \ &\max_{M,\mathbf{P}} \sum^{M}_{i=1}\log_{2}\left(1+\frac{p_{i}y_{i}}{p_{\rm p}z+N_{0}}\right)  \label{Opt_1} \\
&\ \text{s.t.}\ \ \min\left\{{\rm SINR}^{(\rm pr.)}_{j}\right\}^{L}_{j=1}\geq \gamma_{\rm th} \label{Opt_11} \\ 
&\ \ \ \ \ \ \frac{p_{i}y_{i}}{p_{\rm p}z+N_{0}}\geq y_{\rm th} \quad \forall i \label{Opt_111} \\
&\ \ \ \ \ \ \sum^{M}_{i=1}p_{i} \leq p_{\max} \label{Opt_1111}.
\end{align}
\end{subequations}
In the above problem, $\min\{{\rm SINR}^{(\rm pr.)}_{j}\}^{L}_{j=1}$ and $\gamma_{\rm th}$ denote the minimum SINR value at PR (out of $L$ involved PRs) and lower SINR threshold of the primary system, respectively. Also, $y_{\rm th}$ is the lower bound of SINR for the secondary system in order to achieve the required communication quality, while $p_{\max}$ denotes the maximum achievable total power at ST. 

Notably, the problem $\mathcal{P}_{1}$ is an overoptimistic condition and rather infeasible for practical applications since the exact (instantaneous) values of ${\rm SINR}^{(\rm pr.)}$ and $z$ cannot be captured by the secondary system. Since only statistical CSI of secondary-to-primary links (and vice-versa given that channel reciprocity is assumed) is available, we formulate the suboptimal yet realistic version of $\mathcal{P}_{1}$ as
\begin{subequations}
\begin{align}
\mathcal{P}_{2}: \ &\max_{M,\mathbf{P}} \sum^{M}_{i=1}\log_{2}\left(1+\frac{p_{i}y_{i}}{p_{\rm p}\mathbb{E}[z]+N_{0}}\right)  \label{Opt_2} \\
&\ \text{s.t.}\ \ {\rm Pr}\left[\min\left\{{\rm SINR}^{(\rm pr.)}_{j}\right\}^{L}_{j=1}\leq \gamma_{\rm th}\right]\leq \epsilon \label{Opt_22} \\ 
&\ \ \ \ \ \ \frac{p_{i}y_{i}}{p_{\rm p}\mathbb{E}[z]+N_{0}}\geq y_{\rm th} \quad \forall i \label{Opt_222} \\
&\ \ \ \ \ \ \sum^{M}_{i=1}p_{i} \leq p_{\max} \label{Opt_2222},
\end{align}
\end{subequations}
where $\epsilon$ is the upper bound on outage probability that can be supported from the primary system. Note that the knowledge of $\epsilon$ requires negligible computational cost, whereas it can be obtained in the same basis as the channel statistics between the secondary-to-primary links. However, $\mathcal{P}_{2}$ is a non-convex (joint) optimization problem because the objective function is non-convex with respect to both $M$ and $\{p_{i}\}^{M}_{i=1}$. Therefore, in what follows, we propose to solve $\mathcal{P}_{2}$ with respect to $\mathbf{P}=\{p_{i}\}^{M}_{i=1}$ first (given a fixed $M$) and then derive the effective (integer) value of $M$. Doing so, $\mathcal{P}_{2}$ becomes
\begin{subequations}
\begin{align}
\mathcal{P}_{3}: \ &\max_{p_{1},\ldots,p_{M}} \sum^{M}_{i=1}\log_{2}\left(1+\frac{p_{i}y_{i}}{p_{\rm p}\mathbb{E}[z]+N_{0}}\right)  \label{Opt_3} \\
&\ \text{s.t.}\ \ P^{(\rm pr.)}_{{\rm out},\min}(\gamma_{\rm th})\leq \epsilon \label{Opt_33} \\ 
&\ \ \ \ \ \ \frac{p_{i}y_{i}}{p_{\rm p}\mathbb{E}[z]+N_{0}}\geq y_{\rm th} \quad \forall i \label{Opt_333} \\
&\ \ \ \ \ \ \sum^{M}_{i=1}p_{i} \leq p_{\max} \label{Opt_3333},
\end{align}
\end{subequations}
where $P^{(\rm pr.)}_{{\rm out},\min}(\gamma_{\rm th})\triangleq {\rm Pr}[\min\{{\rm SINR}^{(\rm pr.)}_{j}\}^{L}_{j=1}\leq \gamma_{\rm th}]$. It is noteworthy that the objective function in \eqref{Opt_3} represents a lower bound of \eqref{Opt_1}, while the constraint \eqref{Opt_111} is lower bounded by \eqref{Opt_333} (with respect to $z$).\footnote{This occurs by invoking the Jensen's inequality in \eqref{Opt_1} along with the convexity of $\log_{2}(1+p_{i}y_{i}/(p_{\rm p}z+N_{0}))$ on $z$. In a similar basis, the constraint \eqref{Opt_111} is convex with respect to $z$; hence, its lower bound is given by \eqref{Opt_333}.} Thus, solving $\mathcal{P}_{3}$ with respect to $\{p_{i}\}^{M}_{i=1}$ represents an extreme (worst case) scenario of the original problem in $\mathcal{P}_{1}$. This unavoidable cost arises due to the lack of knowledge of the instantaneous secondary-to-primary channel gains; yet, it reflects to a realistic condition for most practical applications.

In $\mathcal{P}_{3}$, the objective function is concave, while all the constraints are linear in $p_{i}$, except \eqref{Opt_33}. To reveal the hidden convexity of the latter constraint, the following lemma is introduced.

\begin{lemma}
\label{Lemma1}
In the case when the average channel gains between $\rm ST$ and $L$ $\rm PRs$ are identical, i.e., $\{\mathbb{E}[h^{(j)}]\}^{L}_{j=1}\triangleq \mathbb{E}[h]$, the outage probability of the minimum SINR at PRs yields as
\begin{align}
P^{(\rm pr.)}_{{\rm out},\min}(\gamma_{\rm th})=1-\left(\frac{p_{\rm p}\mathbb{E}[x]}{p_{\rm p}\mathbb{E}[x]+p_{i}\mathbb{E}[h]\gamma_{\rm th}}\right)^{M L}\exp\left(-\frac{L N_{0}\gamma_{\rm th}}{p_{\rm p}\mathbb{E}[x]}\right).
\label{pout_primary}
\end{align} 
\end{lemma}

\begin{IEEEproof}
The proof is relegated in Appendix~\ref{app_pout_primary}.
\end{IEEEproof}

Since $\partial P^{(\rm pr.)}_{{\rm out},\min}(\gamma_{\rm th})/\partial p_{i}>0$, it is obvious that increasing $p_{i}$, \eqref{pout_primary} also increases. Similarly, increasing $p_{i}$ reflects to a proportional increase of the desired objective function \eqref{Opt_3}. Thus, the optimized solution is to turn the inequality of \eqref{Opt_33} into equality, such that
\begin{align}
\nonumber
&P^{(\rm pr.)}_{{\rm out},\min}(\gamma_{\rm th})=\epsilon\Leftrightarrow 1-\left(\frac{p_{\rm p}\mathbb{E}[x]}{p_{\rm p}\mathbb{E}[x]+p_{i}\mathbb{E}[h]\gamma_{\rm th}}\right)^{M L}\exp\left(-\frac{L N_{0}\gamma_{\rm th}}{p_{\rm p}\mathbb{E}[x]}\right)=\epsilon\\
&\Leftrightarrow p_{i}=\frac{p_{\rm p}\mathbb{E}[x]}{\mathbb{E}[h]\gamma_{\rm th}}\left[\left(\frac{(1-\epsilon)^{\frac{1}{L}}}{\exp\left(-\frac{N_{0}\gamma_{\rm th}}{p_{\rm p}\mathbb{E}[x]}\right)}\right)^{-\frac{1}{M}}-1\right].
\label{ps_constraint_new}
\end{align}

Capitalizing on the above results, $\mathcal{P}_{3}$ recasts as
\begin{subequations}
\begin{align}
\mathcal{P}_{4}: \ &\max_{p_{1},\ldots,p_{M}} \sum^{M}_{i=1}\log_{2}\left(1+\frac{p_{i}y_{i}}{p_{\rm p}\mathbb{E}[z]+N_{0}}\right)  \label{Opt_4} \\
&\ \text{s.t.}\ \ p_{i}\geq \frac{y_{\rm th}\left(p_{\rm p}\mathbb{E}[z]+N_{0}\right)}{y_{i}} \quad \forall i \label{Opt_44} \\
&\ \ \ \ \ \ \sum^{M}_{i=1}p_{i} = \widetilde{p_{\max}} \label{Opt_444},
\end{align}
\end{subequations}
where
\begin{align}
\widetilde{p_{\max}}\triangleq \min\left\{p_{\max},\frac{\left(\frac{(1-\epsilon)^{\frac{1}{L}}}{\exp\left(-\frac{N_{0}\gamma_{\rm th}}{p_{\rm p}\mathbb{E}[x]}\right)}\right)^{-\frac{1}{M}}-1}{\left(\frac{\mathbb{E}[h]\gamma_{\rm th}}{M p_{\rm p}\mathbb{E}[x]}\right)}\right\}.
\label{pmaxx}
\end{align}
The optimization problem $\mathcal{P}_{4}$ is similar to the classical power allocation problem in multiple parallel channels, and the water-filling strategy can be leveraged to obtain the optimal solution.

\begin{proposition}
The optimal value of $p_{i}$, namely $p^{\star}_{i}$, is given by
\begin{align}
p^{\star}_{i}=\frac{y_{i}\left[\widetilde{p_{\max}}+\displaystyle \sum^{M}_{i=1}\left(\frac{1}{y_{i}}\right)(p_{\rm p}\mathbb{E}[z]+N_{0})\right]-M (p_{\rm p}\mathbb{E}[z]+N_{0})}{M y_{i}}.
\label{opt_p}
\end{align}
\end{proposition}

\begin{IEEEproof}
The proof is provided in Appendix~\ref{app_opt_p}.
\end{IEEEproof}

Given $p^{\star}_{i}\quad \forall i$ from \eqref{opt_p}, we are now in a position to determine the effective number of secondary Tx antennas, namely $M^{\star}$, by using the following iterative approach:
\begin{enumerate}
	\item Starting with $M$ available secondary Tx antennas, the constraint \eqref{Opt_44} is computed by using \eqref{opt_p}, such that
	\begin{align}
\frac{y_{i}\left[\widetilde{p_{\max}}+\displaystyle \sum^{M}_{i=1}\left(\frac{1}{y_{i}}\right)(p_{\rm p}\mathbb{E}[z]+N_{0})\right]}{M (p_{\rm p}\mathbb{E}[z]+N_{0})}\geq y_{\rm th}+1.
\label{check_iteratively}
\end{align}
\item In the case when \eqref{check_iteratively} is satisfied, $M = M^{\star}$ and the process is terminated. Otherwise, the same procedure is repeated by setting $M = M-1$, until either \eqref{check_iteratively} is satisfied or $M = 0$.
\end{enumerate}
For completeness of exposition, the proposed iterative approach is formalized in Algorithm~1. At this point is worthy to note that the proposed iterative approach is computationally efficient since $M$ ranges within integer-only values, while \eqref{check_iteratively} is a simple function of the given channel gains.

\begin{algorithm}[t]
	\caption{Effective \# of Secondary Tx Antennas}
	\begin{algorithmic}[1]
		 \INPUT{$M$, $L$, $\{y_{i}\}^{M}_{i=1}$, $y_{\rm th}$, $\gamma_{\rm th}$, $p_{\rm p}$, $\mathbb{E}[z]$, $\mathbb{E}[h]$, $\mathbb{E}[x]$, $N_{0}$, $\epsilon$}
		 \OUTPUT{$M^{\star}$ (the effective number of secondary Tx antennas)}
		  \WHILE{$M>0$}
			\STATE{Compute the inequality \eqref{check_iteratively}, given $M$}
			\IF{\eqref{check_iteratively} is satisfied}
				\STATE $M^{\star} = M$;
				\STATE End of the algorithm;
			\ELSE $\:\:M=M-1$
				\STATE Go to Step 2;
			\ENDIF
		\ENDWHILE 
	\end{algorithmic}
\end{algorithm}

\subsection{Engineering insights}
This section ends with some useful outcomes from the engineering perspective. Firstly, notice that the derived solutions of $\{p^{\star}_{i},M^{\star}\}$ hold for arbitrary antenna arrays. The computational complexity for implementing Algorithm~1 is linear and at most $\mathcal{O}(M)$, which is rather acceptable, especially for conventional MIMO systems (e.g., $4\times 4$ or $8\times 8$ transmission modes).

Secondly, this analytical approach is based on the assumption that the effective channel gain of the primary system (i.e., the numerator in the first equality of \eqref{primarysingle_def}) is exponentially distributed \cite{j:KahlonYanikomeroglu2012}. Nonetheless, this statistical scenario can also be captured from a point-to-point MIMO primary transceiver, in the case when ZF detection is applied and the primary nodes have an equal number of antennas. Doing so, the resultant PDF of the effective channel gain for the primary system is also exponential (e.g., setting $N=M$ in \eqref{ydistr} yields an exponential PDF). 

Thirdly, SINR functions are monotonically increasing functions with respect to the effective channel gain. Also, recall that line-of-sight (LOS) and/or near LOS channel fading conditions are sufficiently modeled by Rician-$K$ or Nakagami-$m$ distributions with $K>0$ and $m>1$, respectively. Moreover, bear in mind that Rayleigh fading (i.e., when $K=0$ for Rician-$K$ or $m=1$ for Nakagami-$m$ models) corresponds to the worst-case scenario in terms of the effective channel gain. Therefore, the optimized solutions presented here can serve as a lower performance bound with regards to the primary system when LOS or near LOS channels are present.

Lastly, the analysis is based on identical $\rm ST-PR$ links, c.f., \eqref{pout_primary_def}. In the case when a single-antenna PT is considered (as in Fig.~\ref{fig1}), only a single PR (out of $L$ PRs) can be served at a given time instance. This case is modeled by setting $L=1$ in \eqref{pmaxx}. On the other hand, in the aforementioned case of a point-point MIMO primary system with $L$ co-located antennas at each node, the assumption of identical $\rm ST-PR$ links seems reasonable since the distance between each secondary Tx antenna is much less than the distance between any secondary Tx and primary Rx antenna.

\subsection{Ergodic capacity of the secondary system}
The normalized ergodic capacity (in bps/Hz) of the $i^{\rm th}$ secondary stream is given by
\begin{align}
C_{i}\triangleq {\rm log}_{2}\left(1+\frac{p^{\star}_{i}y_{i}}{p_{\rm p}z+N_{0}}\right).
\label{cap_def}
\end{align}
The exact value of $C_{i}$ cannot be obtained due to the involvement of $z$, which is unknown. Also, not even the average capacity $\overline{C}_{i}$ (i.e., $\overline{C}_{i}\triangleq \mathbb{E}[C_{i}]$) is tractable for a closed-form solution due to quite a complicated formation of $p_{i}$ in \eqref{opt_p}. Therefore, in what follows, a lower performance bound with respect to $\overline{C}_{i}$ is provided.

\begin{proposition}
A lower bound of the ergodic capacity for the $i^{\rm th}$ secondary stream is expressed as
\begin{align}
C^{(\rm LB)}_{i}={\rm log}_{2}\left(1+\exp(\mathcal{J}_{1}-\mathcal{J}_{2})\right),
\label{cap_lb}
\end{align}
where
\begin{align}
\mathcal{J}_{1}=\frac{\exp\left(-\frac{M^{\star}(p_{\rm p}\mathbb{E}[z]+N_{0})}{\mathbb{E}[y]\widetilde{p_{\max}}}\right)}{\Gamma\left(N-M^{\star}+1,\frac{M^{\star}(p_{\rm p}\mathbb{E}[z]+N_{0})}{\mathbb{E}[y]\widetilde{p_{\max}}}\right)}\sum^{N-M^{\star}}_{k=0}\binom{N-M^{\star}}{k}\frac{k!\left(\frac{M^{\star}(p_{\rm p}\mathbb{E}[z]+N_{0})}{\mathbb{E}[y]\widetilde{p_{\max}}}\right)^{N-M^{\star}-k}}{\left[\psi(k+1)-{\rm ln}\left(\frac{M^{\star}}{\mathbb{E}[y]\widetilde{p_{\max}}}\right)\right]^{-1}},
\label{j1}
\end{align}
and
\begin{align}
\mathcal{J}_{2}={\rm ln}(N_{0})-\exp\left(\frac{N_{0}}{p_{\rm p}\mathbb{E}[z]}\right)\Ei \left(-\frac{N_{0}}{p_{\rm p}\mathbb{E}[z]}\right).
\label{j2}
\end{align}
\end{proposition}

\begin{IEEEproof}
The proof is given in Appendix~\ref{app_c}. 
\end{IEEEproof}

\section{Massive MIMO Energy Harvesting-Enabled Secondary System}
\label{Massive MIMO Energy Harvesting-Enabled Secondary System}
Consider the case when $N\rightarrow +\infty$ with $N\geq M$. Such a scenario corresponds to the so-called massive MIMO transmission. Due to the mode of the proposed operation, there is quite a high possibility that a large antenna array subset could be inactive. This occurs when $M^{\star}\ll M$ and $\{N,M\}$ approach very high values. In turn, there could be $M-M^{\star}$ inactive secondary Tx antennas, which reflect a corresponding high number of available DoF.  

To benefit from the latter DoF, energy harvesting can be implemented onto the secondary Tx, which reflects to the enhancement of the overall power consumption of the considered underlay CR system. For ease of clarity, let $M_{A}$ be the number of active (transmitting) secondary Tx antennas and $M_{I}\triangleq M-M_{A}$ the number of inactive secondary Tx antennas. Then, for a given (fixed) time duration, $\mathcal{T}$, the consumed energy for the secondary system is given by
\begin{align}
E_{C}\triangleq \mathcal{T}\sum^{M_{A}}_{i=1}p^{\star}_{i}.
\label{encons}
\end{align}

\begin{figure}[!t]
\centering
\includegraphics[keepaspectratio,width=2.8in]{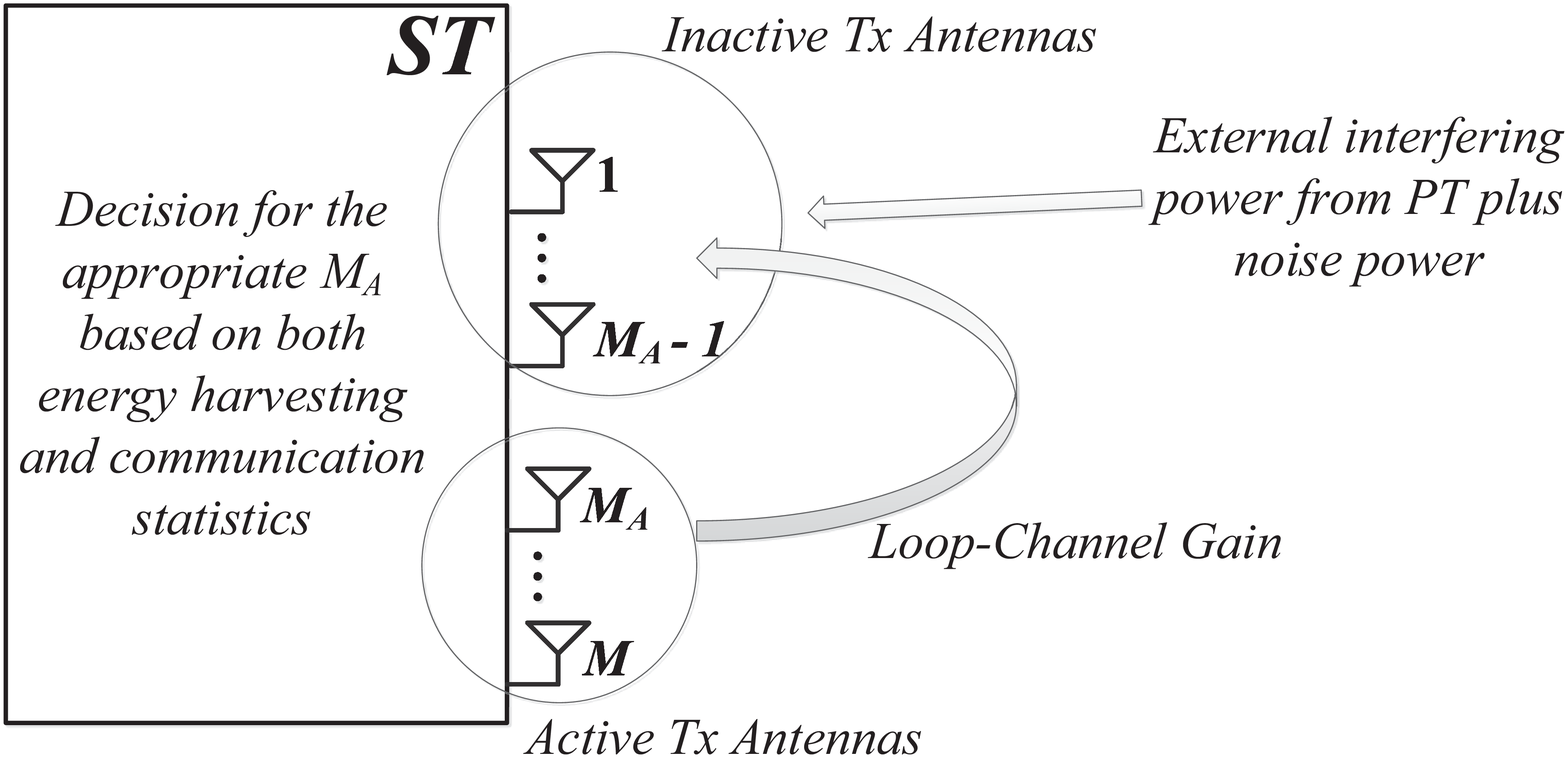}
\caption{The considered system configuration for the energy harvesting-enabled approach.}
\label{fig2}
\end{figure}

Regarding the harvested energy, two popular models have been proposed so far; namely, the linear (e.g., see \cite{j:GuAissa2015,j:DingKrikidis2014} and relevant references therein) and nonlinear \cite{j:BoshkovskaNg2015} models. The latter model is more efficient than the former one since most of the components within an energy harvesting circuit (e.g., diodes, inductors and capacitors) have nonlinear features. Although the nonlinear model captures the true behavior of an energy harvesting circuit quite accurately, it is rather complex and, hence, prohibitive from the performance analysis standpoint. To this end, a simplified piece-wise linear energy harvester model was proposed in \cite{j:DongHossain2016}, which represents an efficient tradeoff between the accuracy and complexity of the aforementioned approaches. According to this model, the harvested energy can be expressed as \cite{j:DongHossain2016}
\begin{align}
E_{H}= \left\{
\begin{array}{ll}
\eta \mathcal{T} P_{H}, &{\rm when }\quad P_{H}<S_{\rm th},\\\\
\eta \mathcal{T} S_{\rm th},& {\rm otherwise},
\end{array}
\right.
\label{enharv}
\end{align} 
where $\eta \in (0,1]$ denotes the energy harvesting efficiency factor, $S_{\rm th}$ is the saturation threshold at the energy harvester, while
\begin{align}
\nonumber
P_{H}&\triangleq \left[\sum^{M_{I}}_{l=1}\sum^{M_{A}}_{i=1}p^{\star}_{i}\left|h^{(L)}_{i,l}\right|^{2}+p_{\rm p}\sum^{M_{I}}_{l=1}\left|h^{\rm (PT-ST)}_{l}\right|^{2}+M_{I}N_{0}\right]\\
&=\Bigg[\sum^{M-M_{A}}_{l=1}\sum^{M_{A}}_{i=1}p^{\star}_{i}\left|h^{(L)}_{i,l}\right|^{2}+p_{\rm p}\sum^{M-M_{A}}_{l=1}\left|h^{\rm (PT-ST)}_{l}\right|^{2}+(M-M_{A})N_{0}\Bigg],
\label{Ph}
\end{align}
where $|h^{(L)}_{i,l}|^{2}$ is the received gain of the \emph{loop-channel} between the $i^{\rm th}$ and $l^{\rm th}$ (co-located) secondary antennas and $|h^{\rm (PT-ST)}_{l}|^{2}\triangleq q_{l}$ (c.f., Fig.~\ref{fig1}) denotes the channel gain between $\rm PT$ and $\rm ST$. The considered system configuration is illustrated in Fig.~\ref{fig2}.

According to the above formulations, reducing $M_{A}$ yields to a proportional increase of the harvested energy at the cost of canceling a corresponding number of transmitted streams, thus, reducing the overall data rate, and vice versa. Hence, the interplay between active and inactive Tx antennas and the derivation of the optimal $M_{A}$, i.e., $M^{\star}_{A}$, is a nontrivial issue. The objective here is the maximization of the achievable sum data rate (given a fixed transmission time interval) of the considered massive MIMO secondary system subject to the previous constraints plus an additional one that satisfies that the ratio of the consumed energy over the harvested energy does not exceed a certain (application-specific) threshold, $P_{\rm th}$. This condition reads as
\begin{subequations}
\begin{align}
\mathcal{P}_{5}: \ &\max_{M_{A},\mathbf{P}} \sum^{M_{A}}_{i=1}\log_{2}\left(1+\frac{p_{i}y_{i}}{p_{\rm p}\mathbb{E}[z]+N_{0}}\right)  \label{Opt_5} \\
&\ \text{s.t.}\ \ p_{i}\geq \frac{y_{\rm th}\left(p_{\rm p}\mathbb{E}[z]+N_{0}\right)}{y_{i}} \quad \forall i \label{Opt_55} \\
&\ \ \ \ \ \ \sum^{M_{A}}_{i=1}p_{i} = \widetilde{p_{\max}} \label{Opt_555}\\
&\ \ \ \ \ \ \frac{E_{C}}{\mathbb{E}\left[E_{H}\right]}\leq P_{\rm th} \label{Opt_5555}.
\end{align}
\end{subequations}
In the last inequality constraint, the expectation operator is due to the fact that (at least) the instantaneous channel gains $q_{l}$ are typically unknown, c.f., \eqref{enharv}. To this end, the consumed energy over the average harvested energy, which both are known (given the corresponding channel gain statistics) should be maintained under a predefined tolerable level $P_{\rm th}$.\footnote{Without loss of generality, we set an upper bound on $P_{\rm th}$, such that $P_{\rm th} \in (0,1]$, so as to prevent the scenario when the consumed energy is higher than the harvested one; thereby, providing energy efficiency.} It turns out that $\mathcal{P}_{5}$ is a non-convex  (joint) optimization problem with respect to $M_{A}$ and $\{p_{i}\}^{M_{A}}_{i=1}$. In what follows, a similar approach as the one presented in the previous Section is adopted; first, the optimal solution of $p_{i}$ is derived given a fixed $M_{A}$, while the optimal value of $M_{A}$ is subsequently obtained.

\subsection{Optimized transmit power $p_{i}$ in the massive MIMO regime given a fixed number of active secondary Tx antennas $M_{A}$}
A key feature of massive MIMO is the \emph{channel hardening} effect, i.e., the small-scale fading tends to average out when asymptotically high antenna arrays are considered. Particularly, according to the law of large numbers, it holds that \cite[Eq. (31)]{j:MatthaiouZFMassiveMIMO2013}
\begin{align}
y_{i}\rightarrow (N-M_{A}+1)\mathbb{E}[y]\quad \forall i,
\label{p_asympt}
\end{align}
where $M_{A}\leq M$.

Capitalizing on the above channel hardening effect, the problem $\mathcal{P}_{5}$, given a fixed $M_{A}$, can be relaxed to
\begin{subequations}
\begin{align}
\mathcal{P}_{6}: \ &\max_{p_{1},\ldots,p_{M_{A}}} \sum^{M_{A}}_{i=1}\log_{2}\left(1+\frac{p_{i}(N-M_{A}+1)\mathbb{E}[y]}{p_{\rm p}\mathbb{E}[z]+N_{0}}\right)  \label{Opt_6} \\
&\ \text{s.t.}\ \ p_{i}\geq \frac{y_{\rm th}\left(p_{\rm p}\mathbb{E}[z]+N_{0}\right)}{(N-M_{A}+1)\mathbb{E}[y]} \quad \forall i \label{Opt_66} \\
&\ \ \ \ \ \ \sum^{M_{A}}_{i=1}p_{i} = \widetilde{p_{\max}} \label{Opt_666}\\
&\ \ \ \ \ \ \frac{E_{C}}{\mathbb{E}\left[E_{H}\right]}\leq P_{\rm th} \label{Opt_6666}.
\end{align}
\end{subequations}
Bearing in mind that the secondary Tx antennas are co-located, while based on the statistical (deterministic) knowledge of the involved parameters in $\mathcal{P}_{6}$, an identical transmit power profile holds for all the secondary Tx antennas in the massive MIMO case. Furthermore, based on \eqref{enharv}, the constraint in \eqref{Opt_6666} can be expressed in terms of $p_{i}$ as
\begin{align}
p_{i}\leq \left\{
\begin{array}{ll}
\frac{\eta P_{\rm th}(M-M_{A})N_{0}+p_{\rm p}\mathbb{E}[q]}{M_{A}-\eta P_{\rm th}\sum^{M-M_{A}}_{l=1}\sum^{M_{A}}_{i=1}\mathbb{E}\left[\left|h^{(L)}_{i,l}\right|^{2}\right]}, &{\rm for}\:\: P_{H}<S_{\rm th},\\\\
\frac{\eta P_{\rm th} S_{\rm th}}{M_{A}},& {\rm otherwise},
\end{array}
\right.
\label{ratio_energy}
\end{align}
where $\mathbb{E}[q]\triangleq \mathbb{E}[q_{l}]\quad \forall l$.

Typically, $|h^{(L)}_{i,l}|$ follows the Rician$-K$ distribution. Its corresponding mean and standard deviation are given, respectively, by $\sqrt{K \Omega_{i,l}/(K+1)}$ and $\sqrt{\Omega_{i,l}/(K+1)}$, where $K$ denotes the Rician$-K$ factor and $\Omega_{i,l}$ is the propagation attenuation of the loop-channel signal between the $i^{\rm th}$ and $l^{\rm th}$ co-located secondary Tx antennas (i.e., the ratio of the received power over the transmit power). Therefore, $\mathbb{E}[|h^{(L)}_{i,l}|^{2}]=\Omega_{i,l}$, whereas $\Omega_{i,l}$ is usually in the order of $-15$dB for co-located antennas \cite{j:ZengZhang2015,j:AtzeniKountouris2017,j:DuarteSabharwal2012}. Also, the Rician$-K$ factor typically takes quite high values, i.e., $K \in [25{\rm dB},40{\rm dB}]$ \cite[Fig.~7]{j:DuarteSabharwal2012}.

\begin{proposition}
In the case when $N\rightarrow +\infty$, the optimized transmit power per antenna for the secondary system approaches
\begin{align}
p^{\star}_{i}\rightarrow \frac{\doublewidetilde{p_{\max}}}{M_{A}}\quad \forall i,
\label{p_asy}
\end{align}
where
\begin{align}
\doublewidetilde{p_{\max}}=\left\{
\begin{array}{ll}
\min\left\{\widetilde{p_{\max}},\frac{\eta M_{A} P_{\rm th}(M-M_{A})N_{0}+p_{\rm p}\mathbb{E}[q]}{M_{A}-\eta P_{\rm th}\sum^{M-M_{A}}_{l=1}\sum^{M_{A}}_{i=1}\Omega_{i,l}}\right\}\triangleq \doublewidetilde{p_{1}},&\\
\ \ \ \ \ \  \ \ \ \ \ \ \ \ \ \ \ \ \ \ \ \ \ \ \ \ \ \ \ \ \ \ \ \ \ \ \ \ \ {\rm when}\:\: P_{H}<S_{\rm th},&\\\\
\min\{\widetilde{p_{\max}},\eta P_{\rm th} S_{\rm th}\}\triangleq \doublewidetilde{p_{2}},\quad \quad \quad {\rm otherwise}.& 
\end{array}
\right.
\label{pmaxxspecial}
\end{align}
\end{proposition}

\begin{IEEEproof}
The proof is provided in Appendix~\ref{app_d}.
\end{IEEEproof}

\subsection{Optimized number of active secondary Tx antennas $M_{A}$ in the massive MIMO regime}
We are now in a position to derive the optimal $M_{A}$, i.e., $M^{\star}_{A}$, given a fixed $p^{\star}_{i}\quad \forall i$. The objective here is to satisfy a predetermined transmission quality for all the active secondary streams (i.e., to satisfy \eqref{Opt_66}) and, at the same time, to preserve the required energy efficiency level (i.e., \eqref{Opt_6666}). According to \eqref{pmaxxspecial}, the following condition should be satisfied
\begin{align}
\doublewidetilde{p_{\max}}&\geq \frac{M_{A} y_{\rm th}\left(p_{\rm p}\mathbb{E}[z]+N_{0}\right)}{(N-M_{A}+1)\mathbb{E}[y]}.
\label{constraintsssnew}
\end{align}

Hence, the optimal $M^{\star}_{A}$ can iteratively be obtained via a simple linear search over the total available number of secondary Tx antennas $M$, such that \eqref{constraintsssnew} holds true. For completeness of exposition, the proposed iterative approach is formalized in Algorithm~2. Notably, the corresponding computational complexity is only $\mathcal{O}(M)$. The proposed iterative approach is computationally efficient since $M$ ranges within integer-only values, while \eqref{constraintsssnew} includes simple functions of the given statistics.

\begin{algorithm}[t]
	\caption{Derivation of $M^{\star}_{A}$}
	\begin{algorithmic}[1]
		 \INPUT{$M$, $N$, $\widetilde{p_{\max}}$, $y_{\rm th}$, $p_{\rm p}$, $\mathbb{E}[z]$, $\mathbb{E}[h]$, $P_{\rm th}$, $\eta$, $N_{0}$, $\mathbb{E}[q_{l}]$, $\mathbb{E}[|h^{(L)}_{i,l}|^{2}]$}
		 \OUTPUT{$M^{\star}_{A}$}
		  \WHILE{$M>0$}
			\STATE{Compute the inequality in \eqref{constraintsssnew}, setting $M_{A}=M$}
			\IF{\eqref{constraintsssnew} is satisfied}
				\STATE $M^{\star}_{A} = M$;
				\STATE End of the algorithm;
			\ELSE $\:\:M=M-1$
				\STATE Go to Step 2;
			\ENDIF
		\ENDWHILE 
	\end{algorithmic}
\end{algorithm}

\subsection{Average capacity of the secondary system}
The average capacity per secondary stream in the massive MIMO case, when $\{N\:\:{\rm and/or}\:\:M\}\rightarrow +\infty$, reads as
\begin{align}
\overline{C}_{i}=\mathbb{E}\left[{\rm log}_{2}\left(1+p^{\star}_{\rm total}\times \frac{(N-M^{\star}_{A}+1)}{p_{\rm p}z+N_{0}}\right)\right],
\label{avcapp}
\end{align} 
where 
\begin{align}
p^{\star}_{\rm total}=\doublewidetilde{p_{1}}{\rm Pr}\left[P_{H}<S_{\rm th}\right]+\doublewidetilde{p_{2}}{\rm Pr}\left[P_{H}\geq S_{\rm th}\right].
\label{ptotal}
\end{align} 
The probability operator ${\rm Pr}\left[P_{H}<S_{\rm th}\right]$ in the above expression denotes the CDF of the harvested energy with respect to $S_{\rm th}$, i.e., $F_{P_{H}}(S_{\rm th})$. Obviously, it represents a key result for the derivation of the average transmit power per each secondary Tx antenna.

\begin{corollary}
The CDF of harvested energy can be expressed in a closed form as
\begin{align}
\nonumber
&F_{P_{H}}(S_{\rm th})=\frac{\left(\frac{m_{K}}{\doublewidetilde{p_{1}}\Omega_{\Sigma}}\right)^{m_{K}}\tilde{S}^{M-M^{\star}_{A}+m_{K}}_{\rm th}}{(p_{\rm p} q)^{M-M^{\star}_{A}}\Gamma(M-M^{\star}_{A}+m_{K}+1)}\\
&\times \Phi_{2}\left(M-M^{\star}_{A},m_{K};M-M^{\star}_{A}+m_{K}+1;\frac{-\tilde{S}_{\rm th}}{p_{\rm p} q},\frac{-m_{K}\tilde{S}_{\rm th}}{\doublewidetilde{p_{1}} \Omega_{\Sigma}}\right),
\label{Proboperator}
\end{align}
where $\tilde{S}_{\rm th}\triangleq S_{\rm th}-(M-M^{\star}_{A})N_{0}$, $m_{K}\triangleq (K+1)^{2}/(2K+1)$ and $\Omega_{\Sigma}\triangleq \sum^{M-M_{A}}_{l=1}\sum^{M_{A}}_{i=1}\Omega_{i,l}$.
\end{corollary}

\begin{IEEEproof}
The proof is relegated in Appendix~\ref{app_e}.
\end{IEEEproof}

Although $\Phi_{2}(\cdot)$ is not yet included as a standard build-in function in most popular mathematical software packages, it can be computed very fast via the efficient algorithm in \cite{Phi2Function}. Moreover, it directly follows that ${\rm Pr}\left[P_{H}\geq S_{\rm th}\right]=1-F_{P_{H}}(S_{\rm th})$.

In the ideal scenario of asymptotically high energy saturation level (i.e., when $S_{\rm th}\rightarrow +\infty$), ${\rm Pr}[P_{H}<S_{\rm th}]=1$ and, thereby, $p^{\star}_{\rm total}=\doublewidetilde{p_{1}}$.\footnote{This occurs due to the fact that the energy harvester behaves linearly when $S_{\rm th}\rightarrow +\infty$, according to the upper case of \eqref{enharv}.} Mathematically, this can be easily verified since \cite[Eq. (3.1)]{j:WaldHenkel2017}
\begin{align}
\nonumber
&\Phi_{2}\left(M-M^{\star}_{A},m_{K};M-M^{\star}_{A}+m_{K}+1;\frac{-\tilde{S}_{\rm th}}{p_{\rm p} q},\frac{-m_{K}\tilde{S}_{\rm th}}{\doublewidetilde{p_{1}} \Omega_{\Sigma}}\right)\\
&\approx \frac{\Gamma(M-M^{\star}_{A}+m_{K}+1)}{\left(\frac{\tilde{S}_{\rm th}}{p_{\rm p} q}\right)^{M-M^{\star}_{A}}\left(\frac{m_{K}\tilde{S}_{\rm th}}{\doublewidetilde{p_{1}} \Omega_{\Sigma}}\right)^{m_{K}}},\quad \tilde{S}_{\rm th}\rightarrow +\infty,
\label{Phiasy}
\end{align}
and, therefore, inserting \eqref{Phiasy} in \eqref{Proboperator} yields $F_{P_{H}}(S_{\rm th})=1$.

Finally, the average channel capacity per each secondary stream is obtained by averaging \eqref{avcapp} over the PDF of $z$.

\begin{proposition}
The average channel capacity per each secondary stream in the massive MIMO regime is derived in a closed form expression as
\begin{align}
\nonumber
\overline{C}_{i}=&{\rm log}_{2}\left(1+\frac{p^{\star}_{\rm total}(N-M^{\star}_{A}+1)}{N_{0}}\right)-\frac{\exp\left(\frac{N_{0}}{p_{\rm p}\mathbb{E}[z]}\right)}{{\rm ln}(2)}\\
&\times \Bigg[{\rm E}_{1}\left(\frac{N_{0}}{p_{\rm p}\mathbb{E}[z]}\right)-\exp\left(\frac{p^{\star}_{\rm total}(N-M^{\star}_{A}+1)}{p_{\rm p}\mathbb{E}[z]}\right){\rm E}_{1}\left(\frac{p^{\star}_{\rm total}(N-M^{\star}_{A}+1)+N_{0}}{p_{\rm p}\mathbb{E}[z]}\right)\Bigg].
\label{avcappcl}
\end{align}
\end{proposition}

\begin{IEEEproof}
The proof is relegated in Appendix~\ref{app_f}.
\end{IEEEproof}

\begin{corollary}
A lower bound on the average channel capacity per each secondary stream in the massive MIMO regime is given by
\begin{align}
\overline{C}_{i}\geq {\rm log}_{2}\left(1+p^{\star}_{\rm total}\times \frac{(N-M^{\star}_{A}+1)}{p_{\rm p}\mathbb{E}[z]+N_{0}}\right).
\label{avcappcllowerbound}
\end{align}
\end{corollary}

\begin{IEEEproof}
Since \eqref{avcapp} is convex on $z$, Jensen's inequality provides the desired result. 
\end{IEEEproof}

\section{Numerical Results}
\label{Numerical Results}
In this section, numerical results are presented and compared with Monte-Carlo (MC) simulation results. In the ensuing simulation experiments, all the involved average channel gains including $\mathbb{E}[x]$, $\mathbb{E}[z]$, $\mathbb{E}[h]$, $\mathbb{E}[q]$, and $\mathbb{E}[y]$ are determined by $(d/d_{\text{ref}})^{-\alpha}$, where $d$ is the Euclidean distance in the unit of meters between two nodes of interest, $d_{\text{ref}}$ is a reference distance of 100 meters (used for normalization), and $\alpha = 4$ denotes the path-loss exponent. Also, the noise variance is normalized, say, $N_{0}=1$, and other parameter setting includes $p_{\max} = 20$dB and $p_{\rm p} = 10$dB with respect to the normalized noise power. Unless otherwise specified, $L=1$, while $\{\gamma_{\rm th},y_{\rm th}\}=1$, which corresponds to a minimum required data rate of $1$bps/Hz for both the primary and secondary services.

\subsection{Non energy harvesting-enabled approach}
\begin{figure}[!t]
\centering
\includegraphics[trim=1.4cm 0.0cm 2.5cm 0cm, clip=true,totalheight=0.4\textheight]{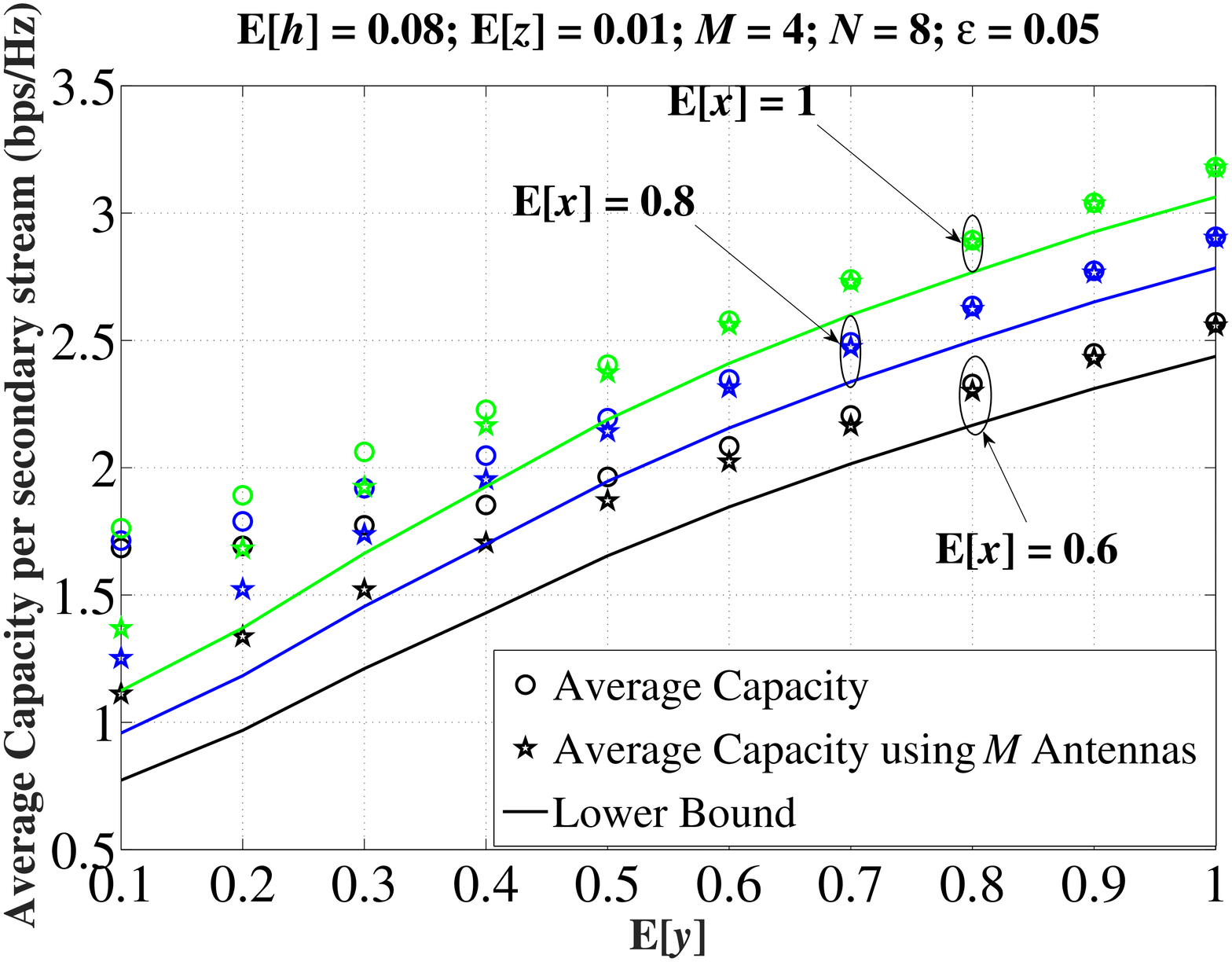}
\caption{The average capacity per secondary stream vs. various values of the effective channel gain between ST and SR.}
\label{fig3}
\end{figure}

\begin{figure}[!t]
\centering
\includegraphics[trim=1.8cm 0.2cm 2.5cm 0cm, clip=true,totalheight=0.4\textheight]{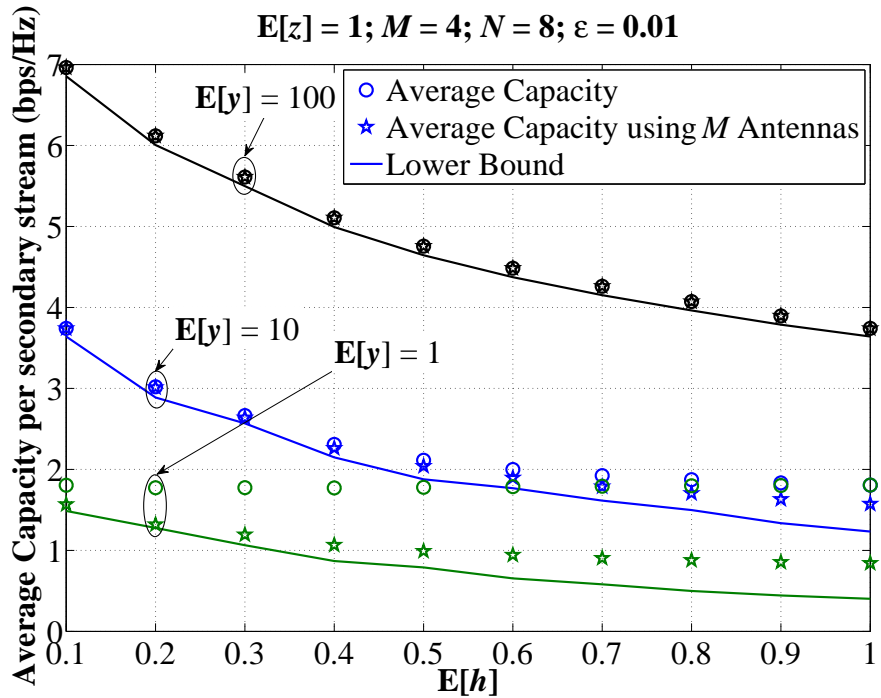}
\caption{The average capacity per secondary stream vs. various values of the interfering channel gain between ST and PRs, where $L=2$.}
\label{fig4}
\end{figure}

\begin{figure}[!t]
\centering
\includegraphics[trim=1.8cm 0.2cm 2.5cm 0cm, clip=true,totalheight=0.4\textheight]{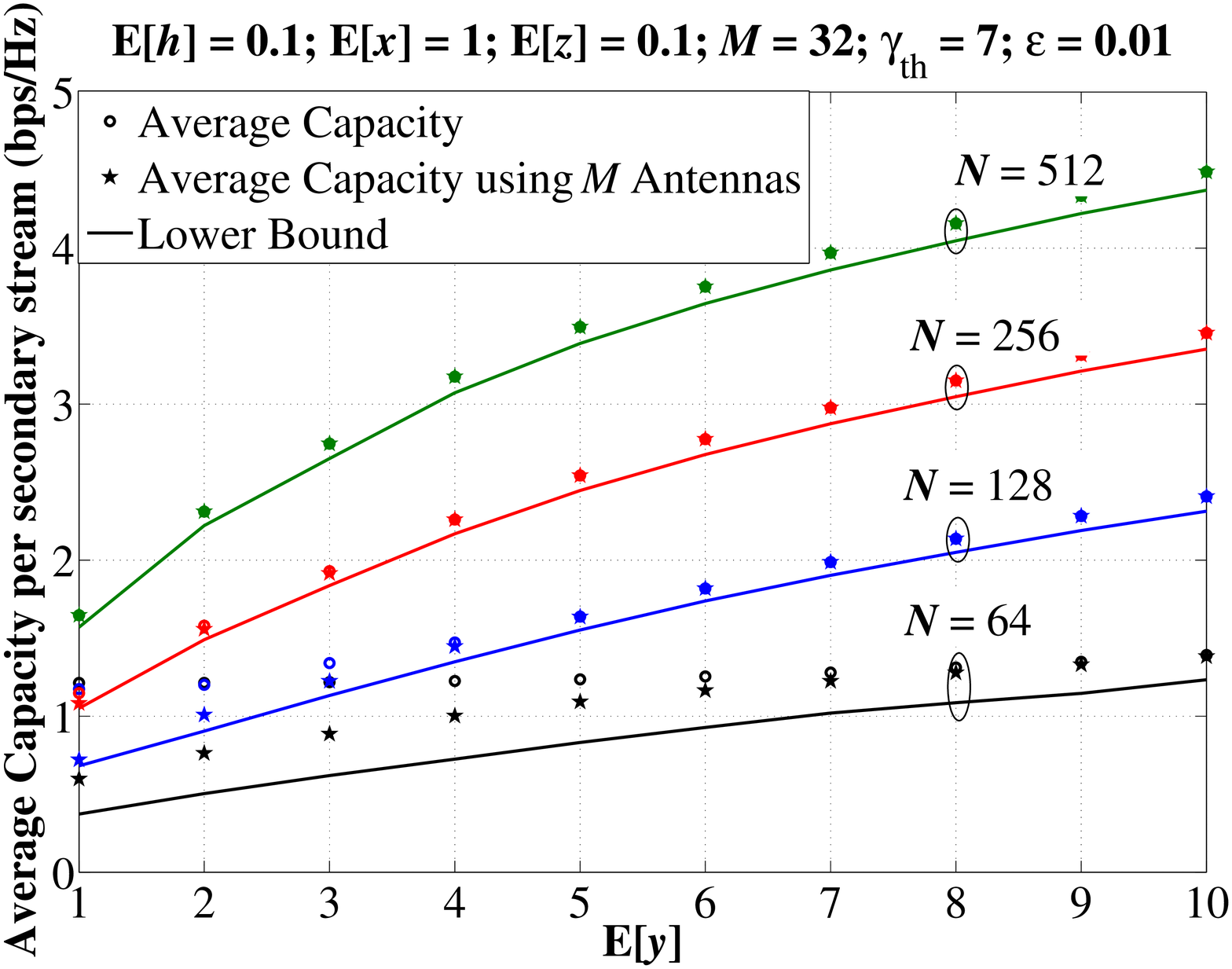}
\caption{The average capacity per secondary stream for a massive MIMO secondary system vs. various values of the effective channel gain between ST and SR.}
\label{fig5}
\end{figure}

Figures~\ref{fig3}-\ref{fig5} illustrate the performance of the proposed approach in terms of the average channel capacity per secondary stream under different system configurations. The lower bound performance curves are labeled as `\emph{Lower Bound}', given by \eqref{cap_lb}; the simulated (exact) average capacity results of the conventional scheme (utilizing all the $M$ Tx secondary antennas) are labeled as `\emph{Average Capacity using $M$ Antennas}'; and the simulated (exact) average capacity results of the proposed scheme (Algorithm~1) are labeled as `\emph{Average Capacity}'.

In Fig.~\ref{fig3}, the average capacity of a secondary MIMO $4\times 8$ underlay CR system is presented. Notably, its performance is enhanced for higher average channel gains of both the primary and secondary systems, $\mathbb{E}[x]$ and $\mathbb{E}[y]$, respectively. This is a reasonable outcome since the total received SINR of both systems increases for closer link-distances (i.e., higher average channel gains). Moreover, the proposed approach outperforms the conventional underlay CR scheme, especially in lower $\mathbb{E}[y]$ values. Essentially, this occurs for two reasons. Firstly, the intra-stream interference (between simultaneously transmitting secondary streams) is higher for $M$ concurrent transmissions rather than $M^{\star}\leq M$ of the proposed scheme. This effect acts more drastically in far-distant ${\rm ST}-{\rm SR}$ links, where the received SINR is proportionally reduced. Secondly, the proposed scheme (i.e., Algorithm~1) drops active secondary Tx antennas, in this case, so as to maintain the acceptable transmission quality as per the constraint in \eqref{check_iteratively}. Doing so, the total received SINR (given $M^{\star}$ transmitted streams) is higher than the SINR of the conventional scheme (given $M$ streams), which in turn reflects to a better channel capacity performance. This effect is more emphatic in bad channel conditions (e.g., lower  $\mathbb{E}[y]$), as illustrated in Fig.~\ref{fig3}. Furthermore, the lower bound is more tight for higher $\mathbb{E}[y]$ values. This occurs because the inequalities in \eqref{app_cc} become tighter for increasing values of $\mathbb{E}[y]$ and vice versa.

The latter observations are further verified in Fig.~\ref{fig4}, where the ${\rm ST}-{\rm SR}$ links are placed within a closer vicinity. For instance, $\mathbb{E}[y]=100$ corresponds to a distance of approximately $30$m. Such link-distances can be realized in dense (urban) networking deployments, as in various heterogeneous cellular infrastructures (e.g., in the range of pico- and/or femto-cell coverage scenarios). The average capacity is reduced for higher interfering power between the secondary Tx and primary Rx (i.e., higher $\mathbb{E}[h]$), as expected. This occurs due to the fact that the secondary transmit power is lower, in this case, so as not to cause unexpected harmful interference to the primary system (satisfying a maximum allowable outage probability of $\epsilon=1\%$). Again, such a transmit power reduction reflects to a proportional reduction on the channel capacity of the secondary system. 

Figure~\ref{fig5} presents the channel capacity performance in the massive MIMO scale. For increasing values of the secondary Rx antennas, $N$, the capacity per secondary stream is enhanced, as it should be. Also, the proposed scheme outperforms the conventional one, especially for worse channel gain conditions with respect the secondary system and/or for lower $N$ values. A noteworthy observation from Fig.~\ref{fig5} (and from Fig.~\ref{fig4}) is that the proposed scheme always satisfies the given transmission quality (i.e., $y_{\rm th}\geq 1$ as per \eqref{check_iteratively}) in contrast to the conventional scheme. In order to do so, the proposed scheme selects $M^{\star}<M$ active secondary Tx antennas to enhance the total received SINR, as previously stated.

\begin{figure}[!t]
\centering
\includegraphics[trim=2.0cm 1.0cm 2.5cm 0cm, clip=true,totalheight=0.4\textheight]{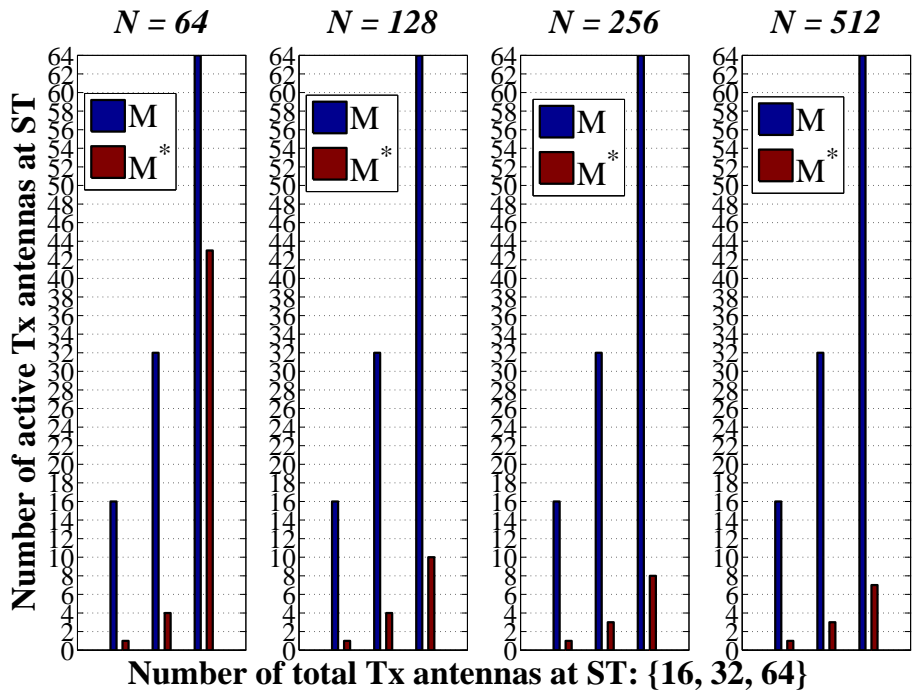}
\caption{Comparison of the conventional massive MIMO CR system and the proposed scheme (Algorithm~1) for various antenna array configurations. Also, $\{\gamma_{\rm th},y_{\rm th}\}=7$, $\{\mathbb{E}[x],\mathbb{E}[y],\mathbb{E}[h],\mathbb{E}[z]\}=1$ and $\epsilon=0.01$.}
\label{fig6}
\end{figure}

To further illustrate the performance of the considered scheme, Fig.~\ref{fig6} compares the number of active secondary Tx antennas used by the conventional and proposed schemes for different system configurations. Although the average capacity per secondary stream is enhanced for higher $N$ values (as demonstrated in Fig.~\ref{fig5}), the number of active secondary Tx antennas ($M^{\star}$) is reduced at the same time. This occurs because the transmit power per secondary antenna, $p^{\star}_{i}$, takes asymptotically very low values as $N\rightarrow +\infty$. Thereby, in order to satisfy a predetermined transmission quality and to enhance the total received SINR at the secondary receiver, the number of active Tx antennas should be reduced. Obviously, there could be a large number of inactive (idle) Tx antennas in massive MIMO setups in a given transmission time interval. This observation stands as our primary motivation to enable the energy harvesting mode-of-operation at the secondary transmitter in order to benefit from a (potentially) high number of inactive Tx antennas.    

\subsection{Energy harvesting-enabled approach}
In the following numerical results, the conventional scheme is labeled as `$M$'; the non energy harvesting-enabled scheme is labeled as `\emph{Algorithm~1}'; while the energy harvesting-enabled scheme is labeled as `\emph{Algorithm~2}'. Also, we set $P_{\rm th}=1$, which represents the case when the consumed energy should not exceed the corresponding harvested energy at a given transmission time interval. For the near-field loop-channel attenuation factor, $\Omega_{i,l}=-15$dB $\forall i,l$ is used, the Rician$-K$ factor is $25$dB, while $\eta=0.85$ \cite{j:Brown1984}. In addition, the normalized (with respect to the AWGN power) saturation threshold at the energy harvester is set to be $S_{\rm th}=100$dB.

\begin{figure}[!t]
\centering
\includegraphics[trim=2.0cm 1.0cm 2.5cm 0cm, clip=true,totalheight=0.4\textheight]{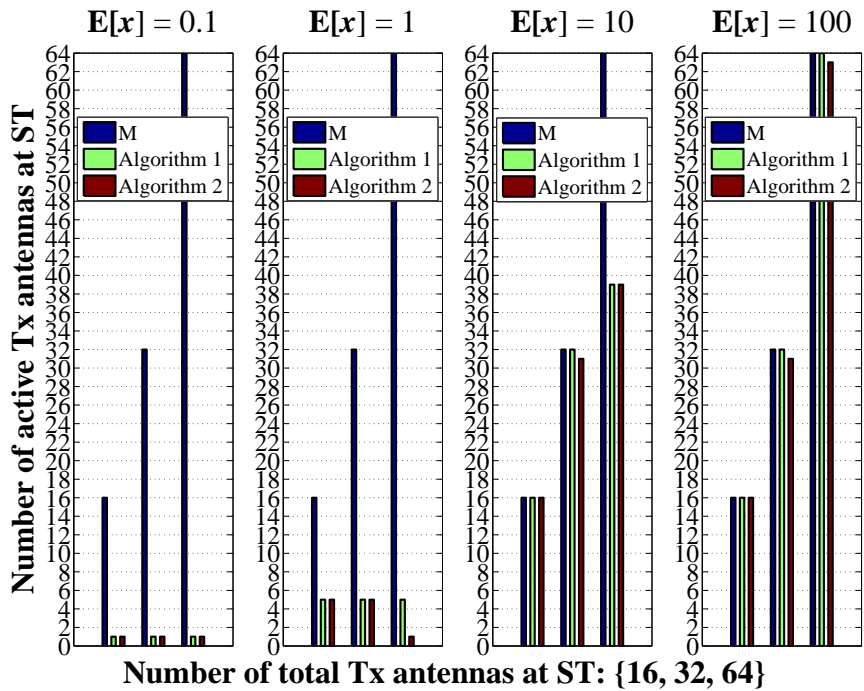}
\caption{Comparison of the conventional massive MIMO CR system, the proposed non energy harvesting-enabled scheme (Algorithm~1) and the proposed energy harvesting-enabled scheme (Algorithm~2) for various antenna array configurations, where $N=128$. Also, $\gamma_{\rm th}=7,y_{\rm th}=3$, $\{\mathbb{E}[h],\mathbb{E}[q],\mathbb{E}[z]\}=1$, $\mathbb{E}[y]=100$, and $\epsilon=0.01$.}
\label{fig7}
\end{figure}

\begin{figure}[!t]
\centering
\includegraphics[trim=2.0cm 1.0cm 2.5cm 0cm, clip=true,totalheight=0.4\textheight]{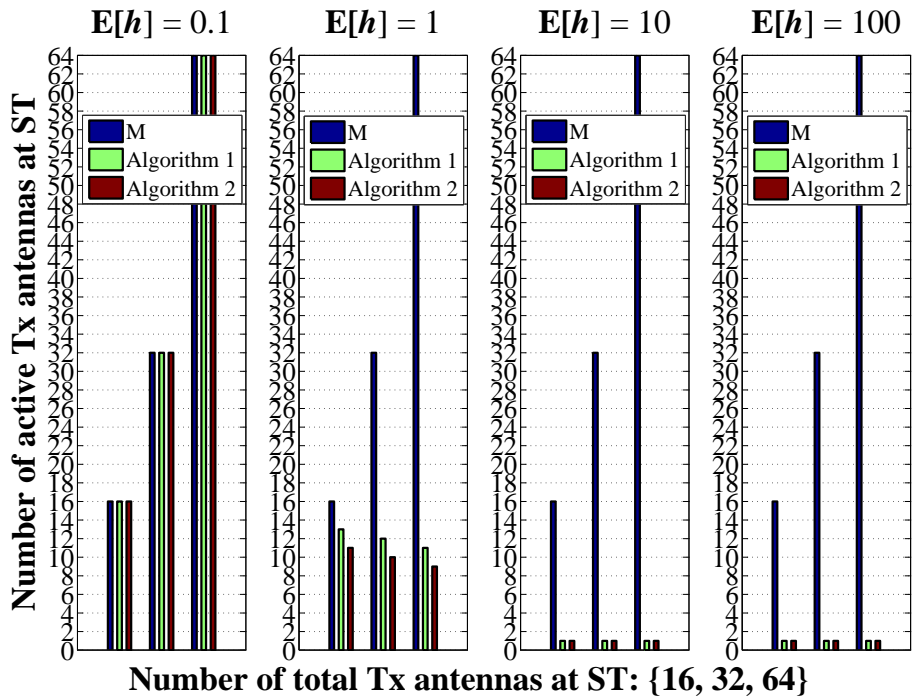}
\caption{Comparison of the conventional massive MIMO CR system, the proposed non energy harvesting-enabled scheme (Algorithm~1) and the proposed energy harvesting-enabled scheme (Algorithm~2) for various antenna array configurations, where $N=128$. Also, $\{\gamma_{\rm th},y_{\rm th}\}=3$, $\{\mathbb{E}[x],\mathbb{E}[q],\mathbb{E}[z]\}=1$, $\mathbb{E}[y]=100$, and $\epsilon=0.01$.}
\label{fig8}
\end{figure}

\begin{figure}[!t]
\centering
\includegraphics[trim=1.8cm 0.2cm 2.5cm 0cm, clip=true,totalheight=0.4\textheight]{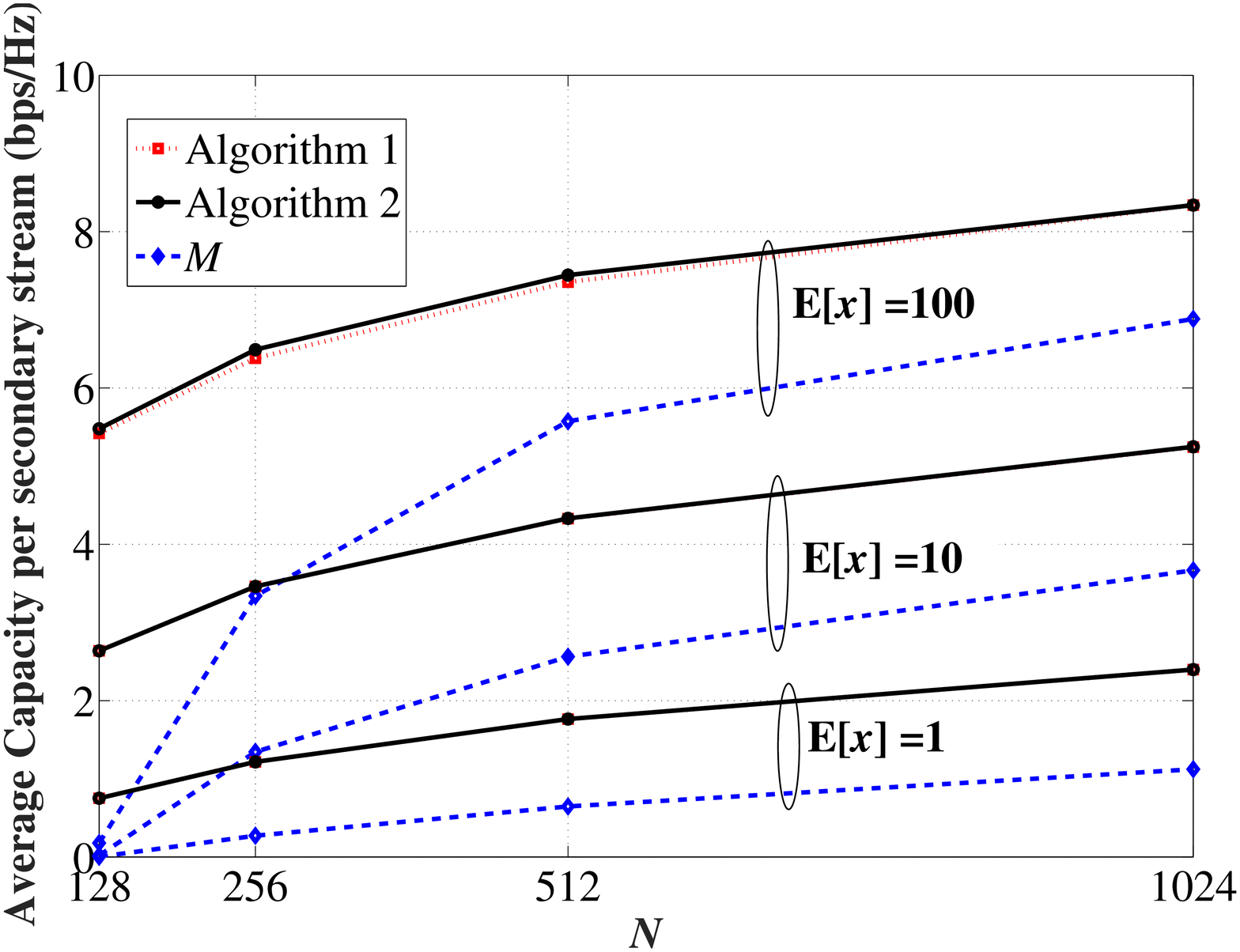}
\caption{The average capacity per secondary stream for a massive MIMO secondary system vs. various values of the secondary Rx antennas $N$, where $M=128$. Also, $\gamma_{\rm th}=7,y_{\rm th}=1$, $\{\mathbb{E}[q],\mathbb{E}[z],\mathbb{E}[y],\mathbb{E}[h]\}=1$, and $\epsilon=0.01$.}
\label{fig9}
\end{figure}

In Figs. \ref{fig7}-\ref{fig9}, the performance of the conventional and the proposed schemes is cross-compared. In particular, Figs. \ref{fig7} and \ref{fig8} indicate the number of active secondary Tx antennas for various system configurations in the massive MIMO scale. Obviously, the number of active antennas is higher for better channel conditions between the primary and secondary systems (i.e., higher and lower values of $\mathbb{E}[x]$ and $\mathbb{E}[h]$, respectively). As previously mentioned, this condition occurs so as to maintain the appropriate communication quality for the secondary system and, at the same time, to preserve the given upper bound on outage probability (i.e., $\epsilon=1\%$) regarding the primary service. A noteworthy observation is that the number of active secondary Tx antennas are greatly reduced in comparison to the conventional (fixed) scenario of $M$ antennas in the case when bad channel conditions are realized. This result emphatically demonstrates that the conventional underlay CR system may reflect to a rather inefficient transmission, a major energy inefficiency, and an increased risk of producing unexpected interference onto the primary system. Interestingly, the performance of both the proposed approaches tend to coincide in most cases; hence, the adoption of energy harvesting-enabled transmission is \emph{highly recommended} in the massive MIMO scale.

Finally, the average channel capacity for various antenna array configurations is presented in Fig. \ref{fig9}, which verify the aforementioned results. The proposed schemes greatly outperform the conventional one, while such a performance improvement gets more emphatic in better channel conditions with respect to the primary system (i.e., higher $\mathbb{E}[x]$ values) and/or higher number of the available secondary Rx antennas.

\section{Conclusions}
\label{Conclusions}
The performance of underlay MIMO CR systems was studied in the case when second-order (statistical) CSI between the primary and secondary systems is available. The secondary system operates using the spatial multiplexing transmission mode, by utilizing ZF detection across the multiple received streams. To enhance the secondary communication quality, an optimal power allocation was considered, according to a predetermined constraint of the maximum allowable outage threshold specified by the primary system. Further, the effective number of secondary Tx antennas was derived via a simple and cost-efficient linear algorithm. In addition, the particular scenario of a massive MIMO secondary system was further studied, where the so-called channel hardening effect was leveraged so as to provide more straightforward results. Most importantly, the certain case when the secondary Tx has energy harvesting capabilities was analyzed. Closed-form expressions for a lower bound on the channel capacity of the secondary system with an arbitrary antenna array and for the average channel capacity of the massive MIMO scale were derived. Finally, both the analytical and numerical results demonstrated that the proposed scheme greatly outperforms the conventional underlay CR approach in terms of channel capacity, efficient secondary communication (without causing unexpected interference onto the primary system) and energy efficiency.

\appendix

\subsection{Derivation of Eq.~\eqref{pout_primary}}
\label{app_pout_primary}
\numberwithin{equation}{subsection}
\setcounter{equation}{0}
It holds that 
\begin{align}
P^{(\rm pr.)}_{{\rm out},\min}(\gamma_{\rm th})=1-[1-P^{(\rm pr.)}_{{\rm out}}(\gamma_{\rm th})]^{L},
\label{pout_primary_def}
\end{align}
where $P^{(\rm pr.)}_{{\rm out}}(\cdot)$ is the outage probability for the SINR of each PR. According to \cite[Eq. (4)]{j:KahlonYanikomeroglu2012}, under Rayleigh faded channels, we have that
\begin{align}
P^{(\rm pr.)}_{{\rm out}}(\gamma_{\rm th})={\rm Pr}\left[\frac{p_{\rm p} x_{j}}{\sum^{M}_{i=1}p_{i} h_{i}+N_{0}}\leq \gamma_{\rm th}\right]=1-\left(1-{\rm Pr}\left[\frac{p_{\rm p} x_{j}}{p_{i} h_{i}}\leq \gamma_{\rm th}\right]\right)^{M}\left(1-{\rm Pr}\left[\frac{p_{\rm p} x_{j}}{N_{0}}\leq \gamma_{\rm th}\right]\right).
\label{primarysingle_def}
\end{align}
In the latter expression, the first and second probability operators correspond to the CDF for the ratio of two independent exponential RVs and CDF of an exponential RV, respectively. Hence, it is straightforward to show that \eqref{primarysingle_def} results to the desired expression in \eqref{pout_primary}.

\subsection{Derivation of Eq.~\eqref{opt_p}}
\label{app_opt_p}
\numberwithin{equation}{subsection}
\setcounter{equation}{0}
The problem in $\mathcal{P}_{4}$ is a convex optimization problem with respect to $p_{i}$ yielding a unique optimal solution. Utilizing the Karush-Kuhn-Tucker conditions and after some straightforward manipulations, we get
\begin{align}
p^{\star}_{i}=\left(\lambda-\frac{(p_{\rm p}\mathbb{E}[z]+N_{0})}{\left(\frac{y_{i}}{1+y_{\rm th}}\right)}\right)^{+}+\frac{y_{\rm th}(p_{\rm p}\mathbb{E}[z]+N_{0})}{y_{i}},
\label{pi-opt}
\end{align} 
where $\lambda$ should satisfy
\begin{align}
\nonumber
&\sum^{M}_{i=1}\left[\left(\lambda-\frac{(p_{\rm p}\mathbb{E}[z]+N_{0})}{\left(\frac{y_{i}}{1+y_{\rm th}}\right)}\right)^{+}+\frac{y_{\rm th}(p_{\rm p}\mathbb{E}[z]+N_{0})}{y_{i}}\right]=\widetilde{p_{\max}}\\
\nonumber
&\Leftrightarrow M \lambda - (p_{\rm p}\mathbb{E}[z]+N_{0})(1+y_{\rm th})\displaystyle \sum^{M}_{i=1}\left(\frac{1}{y_{i}}\right)+y_{\rm th}(p_{\rm p}\mathbb{E}[z]+N_{0})\displaystyle \sum^{M}_{i=1}\left(\frac{1}{y_{i}}\right)=\widetilde{p_{\max}}\\
&\Leftrightarrow \lambda = \frac{1}{M}\left[\displaystyle \sum^{M}_{i=1}\left(\frac{1}{y_{i}}\right)(p_{\rm p}\mathbb{E}[z]+N_{0})+\widetilde{p_{\max}}\right].
\label{lambda}
\end{align} 
Finally, inserting \eqref{lambda} in \eqref{pi-opt}, the desired result in \eqref{opt_p} is extracted.

\subsection{Derivation of Eq.~\eqref{cap_lb}}
\label{app_c}
\numberwithin{equation}{subsection}
\setcounter{equation}{0}
According to \eqref{opt_p} and \eqref{cap_def}, it holds that
\begin{align}
\overline{C}_{i}&\geq {\rm log}_{2}\left(1+\frac{\left(\frac{\widetilde{p_{\max}}}{M^{\star}}\right)y_{i}-M^{\star} (p_{\rm p}\mathbb{E}[z]+N_{0})}{p_{\rm p}z+N_{0}}\right)\geq {\rm log}_{2}\left(1+\exp\left(\mathcal{J}_{1}-\mathcal{J}_{2}\right)\right),
\label{app_cc}
\end{align}
where 
\begin{align}
\mathcal{J}_{1}\triangleq \mathbb{E}\left[{\rm ln}\left(\left(\frac{\widetilde{p_{\max}}}{M^{\star}}\right)y_{i}-M^{\star} (p_{\rm p}\mathbb{E}[z]+N_{0})\right)\right],
\label{jj1}
\end{align}
and
\begin{align}
\mathcal{J}_{2}\triangleq \mathbb{E}\left[{\rm ln}\left(p_{\rm p}z+N_{0}\right)\right].
\label{jj2}
\end{align}
In \eqref{app_cc}, the first inequality occurs by omitting the term $\sum^{M}_{i=1}\left(\frac{1}{y_{i}}\right)(p_{\rm p}\mathbb{E}[z]+N_{0})$ from $p_{i}$. Note that in most practical applications, the underlay CR transceiver maintains quite a small distance due to relatively small Tx power (i.e., high $\mathbb{E}[y]$). Thereby, the latter term approaches a very small value, especially when $M^{\star}$ is high, as in massive MIMO systems. Also, the second inequality of \eqref{app_cc} is based on the fact that ${\rm log}_{2}(1+\exp(x))$ is convex in $x$; thus, applying Jensen's inequality, we get ${\rm log}_{2}(1+\exp({\rm ln}(x)))\geq {\rm log}_{2}(1+\exp(\mathbb{E}[{\rm ln}(x)]))$.

For the derivation of $\mathcal{J}_{1}$, and for ease of presentation clarity, let $W\triangleq (\frac{\widetilde{p_{\max}}}{M^{\star}})y_{i}-M^{\star} (p_{\rm p}\mathbb{E}[z]+N_{0})$. Hence, \eqref{jj1} becomes
\begin{align}
\mathbb{E}\left[{\rm ln}(W)\right]=\int^{+\infty}_{0}{\rm ln}(w)f_{W}(w)dw.
\label{jjj1}
\end{align}
The CDF of $W$ is expressed as
\begin{align}
\nonumber
&F_{W}(x)={\rm Pr}[W=0]+{\rm Pr}[W<x|W>0]={\rm Pr}[W=0]+\frac{{\rm Pr}[0<W<x]}{{\rm Pr}[W>0]}\\
&=F_{y_{i}}\left(\frac{M^{\star}(p_{\rm p}\mathbb{E}[z]+N_{0})}{\widetilde{p_{\max}}}\right)+\frac{F_{y_{i}}\left(\frac{M^{\star}x+M^{\star}(p_{\rm p}\mathbb{E}[z]+N_{0})}{\widetilde{p_{\max}}}\right)-F_{y_{i}}\left(\frac{M^{\star}(p_{\rm p}\mathbb{E}[z]+N_{0})}{\widetilde{p_{\max}}}\right)}{1-F_{y_{i}}\left(\frac{M^{\star}(p_{\rm p}\mathbb{E}[z]+N_{0})}{\widetilde{p_{\max}}}\right)}.
\label{cdfw}
\end{align}
Taking the first derivative of the latter expression, the corresponding PDF yields as
\begin{align}
\nonumber
&f_{W}(x)=\frac{\left(\frac{M^{\star}}{\widetilde{p_{\max}}}\right)f_{y_{i}}\left(\frac{M^{\star}x+M^{\star}(p_{\rm p}\mathbb{E}[z]+N_{0})}{\widetilde{p_{\max}}}\right)}{1-F_{y_{i}}\left(\frac{M^{\star}(p_{\rm p}\mathbb{E}[z]+N_{0})}{\widetilde{p_{\max}}}\right)}\\
&=\frac{\exp\left(-\frac{M^{\star}(p_{\rm p}\mathbb{E}[z]+N_{0})}{\mathbb{E}[y]\widetilde{p_{\max}}}\right)}{\Gamma\left(N-M^{\star}+1,\frac{M^{\star}(p_{\rm p}\mathbb{E}[z]+N_{0})}{\mathbb{E}[y]\widetilde{p_{\max}}}\right)}\sum^{N-M^{\star}}_{k=0}\binom{N-M^{\star}}{k}\frac{(p_{\rm p}\mathbb{E}[z]+N_{0})^{N-M^{\star}-k}x^{k}}{\left(\frac{M^{\star}}{\mathbb{E}[y]\widetilde{p_{\max}}}\right)^{-(N-M^{\star}+1)}\exp\left(\frac{M^{\star}x}{\mathbb{E}[y]\widetilde{p_{\max}}}\right)},
\label{pdfw}
\end{align}
where the fact that $y_{i}\overset{\text{d}}=\mathcal{X}^{2}_{2(N-M+1)}$ and the binomial expansion was used in the last equality of \eqref{pdfw}. Then, utilizing \cite[Eq. (4.352.1)]{tables} and using \eqref{jjj1} and \eqref{pdfw} yields \eqref{j1}.

For the derivation of $\mathcal{J}_{2}$, we have $\mathbb{E}[{\rm ln}(p_{\rm p}z+N_{0})]=\int^{+\infty}_{0}{\rm ln}(p_{\rm p}z+N_{0})f_{z}(z)dz$. Then, using \eqref{zdistr} and utilizing \cite[Eq. (4.337.2)]{tables}, \eqref{j2} arises.

\subsection{Derivation of Eq.~\eqref{p_asy}}
\label{app_d}
\numberwithin{equation}{subsection}
\setcounter{equation}{0}
For a fixed $M_{A}$ and using \eqref{ratio_energy}, the optimization problem $\mathcal{P}_{6}$ becomes
\begin{subequations}
\begin{align}
&\max_{p_{1},\ldots,p_{M_{A}}} \sum^{M_{A}}_{i=1}\log_{2}\left(1+\frac{p_{i}(N-M_{A}+1)\mathbb{E}[y]}{p_{\rm p}\mathbb{E}[z]+N_{0}}\right)  \label{Opt_app1} \\
&\ \text{s.t.}\ \ p_{i}\geq \frac{y_{\rm th}\left(p_{\rm p}\mathbb{E}[z]+N_{0}\right)}{(N-M_{A}+1)\mathbb{E}[y]} \quad \forall i \label{Opt_app2} \\
&\ \ \ \ \ \ \sum^{M_{A}}_{i=1}p_{i} =\left\{
\begin{array}{ll}
\doublewidetilde{p_{1}},&\quad{\rm when}\:\: P_{H}<S_{\rm th},\\\\
\doublewidetilde{p_{2}},&\quad {\rm otherwise}, 
\end{array}
\right. \label{Opt_app3}
\end{align}
\end{subequations}
where $\doublewidetilde{p_{1}}$ and $\doublewidetilde{p_{2}}$ are defined back in \eqref{pmaxxspecial}. The above problem tries to maximize the sum-capacity of the secondary system, while maintaining the predefined transmission quality of both primary and secondary systems, at the same time. According to \eqref{Opt_app3}, the energy efficiency of the secondary system is also considered for two distinct cases; when the harvested level is below the energy saturation level (i.e., $P_{H}<S_{\rm th}$) or when it exceeds this level (i.e., $P_{H}\geq S_{\rm th}$).

In essence, the involved statistics in \eqref{Opt_app3} change very slow with respect to their corresponding instantaneous counterparts. Thus, the harvested energy in the $(\mathcal{T}-1)$ transmission time interval can be used to optimally determine $p_{i}$ in the next transmission time interval $\mathcal{T}$, by solving the above problem given $P_{H}[\mathcal{T}-1]$.  

In turn, the above optimization problem is in the form of $\mathcal{P}_{4}$, yielding the optimal solution as per \eqref{opt_p}. Finally, inserting \eqref{p_asympt} in \eqref{opt_p} and substituting $M$ with $M_{A}$ yields \eqref{p_asy}.

\subsection{Derivation of Eq.~\eqref{Proboperator}}
\label{app_e}
\numberwithin{equation}{subsection}
\setcounter{equation}{0}
Let $\mathcal{Z}\triangleq \mathcal{Z}_{1}+\mathcal{Z}_{2}$, with
\begin{align}
\mathcal{Z}_{1}\triangleq \doublewidetilde{p_{1}} \sum^{M-M_{A}}_{l=1}\sum^{M_{A}}_{i=1}\left|h^{(L)}_{i,l}\right|^{2},
\end{align}
and
\begin{align}
\mathcal{Z}_{2}\triangleq p_{\rm p}\sum^{M-M_{A}}_{l=1}\left|h^{\rm (PT-ST)}_{l}\right|^{2}.
\end{align}
Then, according to \eqref{Ph}, we have that 
\begin{align}
{\rm Pr}[P_{H}<S_{\rm th}]=F_{\mathcal{Z}}(\tilde{S}_{\rm th}).
\label{PrZz}
\end{align}
Also, $\mathcal{Z}$ includes the sum of a non-central (i.e., $\mathcal{Z}_{1}$) and central (i.e., $\mathcal{Z}_{2}$) chi-squared RVs, which are mutually independent. 

In general, there are two types of signal addition; namely, coherent and incoherent addition. The former type occurs whenever the carrier frequencies of the individual signals are equal and the random phase fluctuations are small during the transmission time interval. Otherwise, the latter type of incoherent addition occurs \cite{j:yacoub}. In the considered case of closely spaced co-located antennas, coherent signal addition is used to model $\mathcal{Z}_{1}$ having a PDF denoted as \cite[Eq. (15)]{j:yacoub}
\begin{align}
f_{\mathcal{Z}_{1}}(x)=\frac{(K+1)\exp\left(-\frac{(K+1) x}{\doublewidetilde{p_{1}} \Omega_{\Sigma}}\right)}{\doublewidetilde{p_{1}} \Omega_{\Sigma}\exp(K)}I_{0}\left(2\sqrt{K(K+1)\frac{x}{\doublewidetilde{p_{1}} \Omega_{\Sigma}}}\right).
\label{pdfZ1}
\end{align}
In addition, the PDF of $\mathcal{Z}_{2}$ is given by
\begin{align}
f_{\mathcal{Z}_{2}}(x)=\frac{x^{M-M_{A}-1}\exp\left(-\frac{x}{p_{\rm p}q}\right)}{\Gamma(M-M_{A})(p_{\rm p}q)^{M-M_{A}}}.
\label{pdfZ2}
\end{align}
For analytical tractability, we proceed by approximating the PDF in \eqref{pdfZ1} using a gamma distribution. In particular, it is well-known that Rician$-K$ fading can be efficiently approached by the Nakagami$-m$ distribution, by appropriately determining its $m$-parameter \cite[Eq. (2.26)]{simonAlouini2005digital}, such as
\begin{align}
f_{\mathcal{Z}_{1}}(x)=\frac{\left(\frac{m_{K}}{\doublewidetilde{p_{1}} \Omega_{\Sigma}}\right)^{m_{K}}}{\Gamma(m_{K})}\:x^{m_{K}-1}\exp\left(-\frac{m_{K} x}{\doublewidetilde{p_{1}} \Omega_{\Sigma}}\right).
\label{pdfZzz1}
\end{align}
Recall that the Rician$-K$ factor reaches quite high values (i.e., $>25$dB) in the considered case. Doing so, both distributions \eqref{pdfZ1} and \eqref{pdfZzz1} approach a Dirac's delta-like PDF, whereas they perfectly match.

Based on \eqref{pdfZ2} and \eqref{pdfZzz1}, the corresponding MGFs yield as
\begin{align}
M_{\mathcal{Z}_{1}}(s)=\mathbb{E}\left[\exp\left(-s \mathcal{Z}_{1}\right)\right]=\left(\frac{m_{K}}{\doublewidetilde{p_{1}} \Omega_{\Sigma}}\right)^{m_{K}}\left(\frac{m_{K}}{\doublewidetilde{p_{1}} \Omega_{\Sigma}}+s\right)^{-m_{K}},
\label{mgfZzz1}
\end{align}
and
\begin{align}
M_{\mathcal{Z}_{2}}(s)=\mathbb{E}\left[\exp\left(-s \mathcal{Z}_{2}\right)\right]=\left(p_{\rm p} q\right)^{-(M-M_{A})}\left(\frac{1}{p_{\rm p} q}+s\right)^{-(M-M_{A})}.
\label{mgfZ2}
\end{align}

Finally, it holds that
\begin{align}
F_{\mathcal{Z}}(\tilde{S}_{\rm th})=\mathcal{L}^{-1}\left\{s^{-1}M_{\mathcal{Z}_{1}}(s)M_{\mathcal{Z}_{2}}(s);s;\tilde{S}_{\rm th}\right\},
\label{cdfmgf}
\end{align}
which is directly obtained in a closed-form solution with the aid of \cite[Eq. (2.1.3.1)]{prudnikov1992integrals}, thus extracting the desired result.

\subsection{Derivation of Eq.~\eqref{avcappcl}}
\label{app_f}
\numberwithin{equation}{subsection}
\setcounter{equation}{0}
From \eqref{avcapp}, we get
\begin{align}
\nonumber
&\overline{C}_{i}=\int^{+\infty}_{0}{\rm log}_{2}\left(1+\frac{p^{\star}_{\rm total}(N-M^{\star}_{A}+1)}{p_{\rm p}z+N_{0}}\right)\frac{\exp\left(-\frac{z}{\mathbb{E}[z]}\right)}{\mathbb{E}[z]}dz\\
\nonumber
&={\rm log}_{2}\left(1+\frac{p^{\star}_{\rm total}(N-M^{\star}_{A}+1)}{N_{0}}\right)-\frac{1}{{\rm ln}(2)}\\
\nonumber
&\ \ \ \ \times \int^{+\infty}_{0}\left[\frac{p_{\rm p}\:p^{\star}_{\rm total}(N-M^{\star}_{A}+1)}{(p_{\rm p}z+N_{0})^{2}\left(\frac{p^{\star}_{\rm total}(N-M^{\star}_{A}+1)}{p_{\rm p}z+N_{0}}+1\right)}\right]\exp\left(-\frac{z}{\mathbb{E}[z]}\right)dz\\
\nonumber
&={\rm log}_{2}\left(1+\frac{p^{\star}_{\rm total}(N-M^{\star}_{A}+1)}{N_{0}}\right)-\frac{p_{\rm p}}{{\rm ln}(2)}\\
&\times \int^{+\infty}_{0}\left(\frac{1}{p_{\rm p} z+N_{0}}-\frac{1}{p^{\star}_{\rm total}(N-M^{\star}_{A}+1)+p_{\rm p} z+N_{0}}\right)\exp\left(-\frac{z}{\mathbb{E}[z]}\right)dz,
\label{avcapppp}
\end{align} 
where the second equality of \eqref{avcapppp} is obtained by implementing integration by parts. Hence, utilizing \cite[Eq. (3.462.15)]{tables} and after some straightforward manipulations, the desired result is presented in \eqref{avcappcl}.

\bibliographystyle{IEEEtran}
\bibliography{IEEEabrv,References}

\vfill

\end{document}